\documentclass[fleqn,usenatbib]{mnras}

\usepackage{newtxtext,newtxmath}
\usepackage[T1]{fontenc}

\DeclareRobustCommand{\VAN}[3]{#2}
\let\VANthebibliography\thebibliography
\def\thebibliography{\DeclareRobustCommand{\VAN}[3]{##3}\VANthebibliography}

\usepackage{bm}
\usepackage{graphicx}	% Including Fig. files
\usepackage{amsmath}	% Advanced maths commands
\usepackage[dvipsnames]{xcolor}
\usepackage{natbib}
\usepackage{longtable}
\usepackage{upgreek}
\usepackage{ulem}
\usepackage[dvipsnames]{xcolor}
\usepackage{cancel}%For editing, can be removed before the submission
\defcitealias{SPI}{SPI22} %grs: if using this alias, it should be clearly defined in the text (and also noted in the bibliogrpahy)
\newcommand{\dd}{\mathrm{d}}        %differentialhttps://www.overleaf.com/project/5f203812a32b1700019f86e2
\newcommand\deriv[2]{\frac{\partial#1}{\partial#2}}%partial derivative
\newcommand\dderiv[2]{\frac{\text{D}#1}{\text{D}#2}}%Lagrangian derivative
\renewcommand{\vec}{\bm}% vector convention
\newcommand{\cm}{\,{\rm cm}}
\newcommand{\g}{\,{\rm g}}
\newcommand{\K}{\,{\rm K}}
\newcommand{\kms}{\,{\rm km}\,{\rm s}^{-1}}
\newcommand{\km}{\,{\rm km}}
\newcommand{\s}{\,{\rm s}}
\newcommand{\kpc}{\,{\rm kpc}}
\newcommand{\muG}{\,{\upmu\rm G}}

\newcommand{\Gyr}{\,{\rm Gyr}}
\newcommand{\Myr}{\,{\rm Myr}}
\newcommand{\ecr}{\epsilon_\text{cr}}%cosmic ray energy density
\newcommand{\ecrzero}{\epsilon_{\text{cr},0}}%cosmic ray energy density
\newcommand{\cra}{_\text{cr}}%subscript cosmic rays
\newcommand{\crit}{_\text{c}}%subscript critical
\newcommand{\therm}{_\text{th}}%subscript thermal
\newcommand{\m}{_\text{m}}%subscript magnetic 
\newcommand{\kin}{_\text{k}}%subscript kinetic
\newcommand{\sound}{_\text{s}}%subscript magnetic 
\newcommand{\ecri}{\epsilon_{\rm{cr0}}}%imposed cosmic ray energy density

\newcommand{\rms}{{\text{r.m.s.}}}
\newcommand{\const}{{\text{const}}}

\newcommand\mean[1]{\overline{#1}}% average
\newcommand\meanh[1]{\langle#1\rangle_{\text h}}%horizontal average
\newcommand\Meanh[1]{\left\langle#1\right\rangle_{\text h}}%horizontal average, big<>
\newcommand\bh{\tilde{b}}%magnetic deviation from horizontal average
\newcommand\uh{\tilde{u}}%velocity deviation from horizontal average

\newcommand{\SimA}{{\sf $\Omega$00N}}
\newcommand{\SimB}{{\sf $\Omega$30S}}
\newcommand{\SimC}{{\sf $\Omega$60S}}
\newcommand{\SimD}{{\sf $\Omega$30N}}

\definecolor{coralred}{rgb}{0.765625, 0.1171875, 0.2265625}

\definecolor{burntorange}{rgb}{0.8, 0.33, 0.0}
\definecolor{brown}{rgb}{0.63, 0.17, 0.0}%{0.43, 0.17, 0.0}

\definecolor{magenta}{rgb}{1.0,0.0,1.0}

\title[Nonlinear Parker instability and rotation]{Steady states of the Parker instability: the effects of rotation}

\author[Tharakkal et al.]{Devika Tharakkal$^{1}$, Anvar Shukurov$^{1}$,
Frederick A.~Gent$^{2,1}$, Graeme R.~Sarson$^{1}$, Andrew Snodin$^{3,1}$\\
$^{1}$School of Mathematics, Statistics and Physics, Newcastle University, Newcastle upon Tyne, NE1 7RU, UK\\
$^{2}$Astroinformatics, Department of Computer Science, Aalto University, PO Box 15400, FI-00076 Espoo, Finland\\ 
$^{3}$UKAEA, Culham Science Centre, Abingdon, OX14 3DB, UK\\ 
}

\date{Accepted XXX. Received YYY; in original form ZZZ}

\pubyear{2022}

\begin{document}
\label{firstpage}
\pagerange{\pageref{firstpage}--\pageref{lastpage}}
\maketitle

%----------------------------------------
\begin{abstract}
We model the Parker instability in vertically stratified isothermal gas using
non-ideal MHD three-dimensional simulations. Rotation, especially differential,
more strongly and diversely affects the nonlinear state than the linear stage
(where we confirm the most important conclusions of analytical models), and
stronger than any linear analyses predict. Steady state magnetic fields are
stronger and cosmic ray energy density higher than in comparable nonrotating
systems. Transient gas outflows induced by the nonlinear instability persist
longer, of order 2 Gyr, with rotation. Stratification combined with
(differential) rotation drives helical flows, leading to mean-field dynamo.
Consequently, the nonlinear state becomes oscillatory (while both the linear
instability and the dynamo are non-oscillatory). The horizontal magnetic field
near the midplane reverses its direction propagating  to higher altitudes as
the reversed field spreads buoyantly. The spatial pattern of the large-scale
magnetic field may explain the alternating magnetic field directions in the
halo of the edge-on galaxy NGC 4631. Our model is unique in producing a
large-scale magnetic structure similar to such observation. Furthermore, our
simulations show that the mean kinetic helicity of the magnetically driven
flows has the sign opposite to that in the conventional non-magnetic flows.
This has profound consequences for the nature of the dynamo action and
large-scale magnetic field structure in the coronae of spiral galaxies which
remain to be systematically explored and understood. We show that the energy
density of cosmic rays and magnetic field strength are not correlated at scales
of order a kiloparsec.
\end{abstract}

\begin{keywords}
instabilities -- magnetic fields -- MHD -- cosmic rays -- ISM: structure --
galaxies: magnetic fields
\end{keywords}

%%%%%%%%%%%%%%%%%%%%%%%%%%%%%%%%%%%%%%%%%%%%%%%%%%
\section{Introduction}\label{In}
The Parker instability is a magnetic Rayleigh--Taylor or magnetic buoyancy instability modified by cosmic rays that carry negligible weight but exert significant pressure. The instability is an important element of the large-scale dynamics of the interstellar medium (ISM) as it affects the vertical distributions of the gas, magnetic fields and cosmic rays and can drive gas outflows, thereby affecting the star formation. In our previous work \citep{SPI}, we explored the development of the instability, with a focus on its nonlinear saturation, in a non-rotating disc with imposed unstable distributions of the gas, magnetic field and cosmic rays. Among the essentially nonlinear features of the instability are a transient gas outflow in the weakly nonlinear stage and a strong redistribution of magnetic fields, cosmic rays and thermal gas, resulting in a thinner thermal gas disc and very large scale heights and low energy densities of the magnetic field and cosmic rays. In this paper, we address the effect of rotation on the Parker instability. 

Rotation is known to reduce the growth rate of the weak perturbations but it does not suppress the instability completely \citep{ZwKu1975,FogTag1994,FogTag1995,TMatRMat1998,KoHaOt2003}. However, rotation introduces a fundamentally new feature to the system: under the action of the Coriolis force, the gas flows produced by the instability become helical and can drive mean-field dynamo action that generates a magnetic field at a large scale comparable to that of the initial unstable configuration. \citet{Hanasz1997b}, \citet{Hanasz1997a,Hanasz1998} and \citet{The00a} simulate numerically the mean-field dynamo action driven by the magnetic buoyancy with and without cosmic rays, while \citet{MSS99} present an analytical formulation. A striking feature of the nonlinear evolution of a rotating system, noticed by \citet{Machida2013} in their simulations of the galactic dynamo using ideal magnetohydrodynamics (MHD), is the possibility of quasi-periodic magnetic field reversals at the time scale of $1.5\Gyr$, both near the disc midplane and at large altitudes. This appears to be an essentially nonlinear effect that relies on rotation since the linear instability does not develop oscillatory solutions and the nonlinear states are not oscillatory without rotation \citep{SPI}. \citet[][their Section~7.1]{FogTag1994} find that the Parker instability can be oscillatory in a certain range of the azimuthal wave numbers. \citet{Machida2013} relate the reversals to the magnetic flux conservation, but we note that the \textit{large-scale} magnetic flux is not conserved when the mean-field dynamo is active. Our simulations of the nonlinear Parker instability in a rotating system suggest a different, more subtle explanation that relies on the correlations between magnetic and velocity fluctuations not dissimilar to those arising from the $\alpha$-effect that drives the mean-field dynamo action (see below). Large-scale magnetic fields whose horizontal direction alternates with height emerge in the simulations of mean-field dynamo action by \citet{Hanasz2004}. 
This spatial pattern may be related to the field reversals near the midplane.  

We explore the effects of rotation on the Parker instability in a numerical model similar to that of \citet{SPI}, quantifying both its linear and nonlinear stages and identifying the roles of the Coriolis force and the velocity shear of the differential rotation. We consider the instability in a local rectangular box with parameters similar to those of the Solar neighbourhood of the Milky Way. The structure of this paper is as follows. Section~\ref{sec:NM_rot} describes briefly the numerical model, and in Section~\ref{sec:LI_rot} we consider the linear stage of the instability. Section~\ref{sec:NI_rot} presents a detailed comparison of the distributions of the thermal and non-thermal components of the system in the nonlinear, saturated stage of the instability and how they change when the rotational speed and shear rate vary. in Section~\ref{sec:mr}, we clarify the mechanism of the magnetic field reversal and Section~\ref{VFFB} discusses the effects of rotation on the systematic vertical flows. The mean-field dynamo action of the motions induced by the instability is our subject in Section~\ref{sec:tc} where we discuss the kinetic and magnetic helicities.

%--------------------------------------------------------------------------
\begin{table}
\centering
\caption{The list of simulation runs discussed: the numerical resolutions along each axis, the angular velocity and rotational shear, and the instability growth rate computed for $u_z$ and $b_z$.}
\begin{tabular}{@{}ccccc@{}}
\hline
&$(\Delta x,\Delta y,\Delta z)$  &$\Omega$       &$S$                     &$\Gamma$\\
&[pc]     &[km\,s$^{-1}\kpc^{-1}$]   &[km\,s$^{-1}\kpc^{-1}$] &[Gyr$^{-1}]$\\
\hline
\SimA  & (15,7,13)   &\phantom{$3$}0     &\phantom{$-3$}0 &23\\
\SimD  & (31,15,27)  &30                 &\phantom{$-3$}0 &22\\
\SimB  & (31,15,27)  &30                 &$-30$           &12\\
\SimC  & (31,15,27)  &60                 &$-60$      &\phantom{$2$}7\\
\hline
    \end{tabular}    
    \label{tab:sims}
\end{table}
%--------------------------------------------------------------

%----------------------------------------------------
\section{Basic equations and the numerical model}\label{sec:NM_rot}
We use a model very similar to that of \citet{SPI}, with the only difference being that we now consider rotating systems, with either a solid-body or differential rotation. We consider the frame rotating at the angular velocity of the centre of the domain with the $z$-axis aligned with the gravitational acceleration and the angular velocity $\vec{\Omega}$, the $y$-axis directed along the azimuth and the $x$-axis parallel to the radial direction of the local cylindrical frame. Vector $x$-components are occasionally referred to as radial, while $y$-components are called azimuthal.

The non-ideal MHD equations are formulated for the gas density $\rho$, its velocity $\vec{U}$, total pressure $P$ (which includes the thermal, magnetic and cosmic-ray contributions), magnetic field $\vec{B}$ and its vector potential $\vec{A}$, and the energy density of cosmic rays $\ecr$. 
The initial conditions represent an unstable magneto-hydrostatic equilibrium, and the corresponding distributions $\rho_0$, $\vec{B}_0$ and $\ecrzero$ in $z$ are maintained throughout the simulation as a background state. We solve for the deviations from them, denoted $\rho'$ for the density, $\vec{u}$ for the velocity, $P'$ for the total pressure, $\vec{b}$ for the magnetic field and $\vec{a}$ for its vector potential, and $\ecr'$ and $\vec{F}'$ for the cosmic-ray energy density and flux. Cosmic rays are described in the fluid approximation with non-Fickian diffusion, so we have separate equations for their energy density and flux. The governing equations are solved numerically in a rectangular shearing box of the size $4\times 4\times 3.5\kpc^3$ along the $x$, $y$ and $z$ axes, respectively, with the mid-plane at $z=0$ 
and $|z|\leq1.75\kpc$.  
The boundary conditions are periodic in $x$, sliding-periodic in $y$ and allow for a free exchange of matter through the top and bottom of the domain as specified in detail by \citet{SPI}. 

The total velocity is given by $\vec{U} = \vec{U}_0 + \vec{u}$, where $\vec{U}_0 = Sx \hat{\vec{y}}$ is the mean rotation velocity
in the rotating frame
with the shear rate $S=x\,\dd\Omega/\dd x$, and $\vec{u}$ is the deviation from this, associated with the instability. For a solid-body rotation, $S=0$, we have $\vec{U}_0=0$. Both $S$ and $\Omega$ are assumed to be independent of $z$ 
and $S<0$ for realistic galactic rotation profiles.
We neglect the vertical gradient of $\Omega$ and $S$; for its observed magnitude of order $-15\text{--}25\km\kpc^{-1}$ \citep[Section~10.2.3 of][and references therein]{ShSu22}, $\Omega$ and $S$ only vary by about 10--15 per cent within $|z|\lesssim1.5\kpc$.

%-------------------------------------------------
\begin{figure*}
    \centering
    \includegraphics[width=0.9\textwidth]{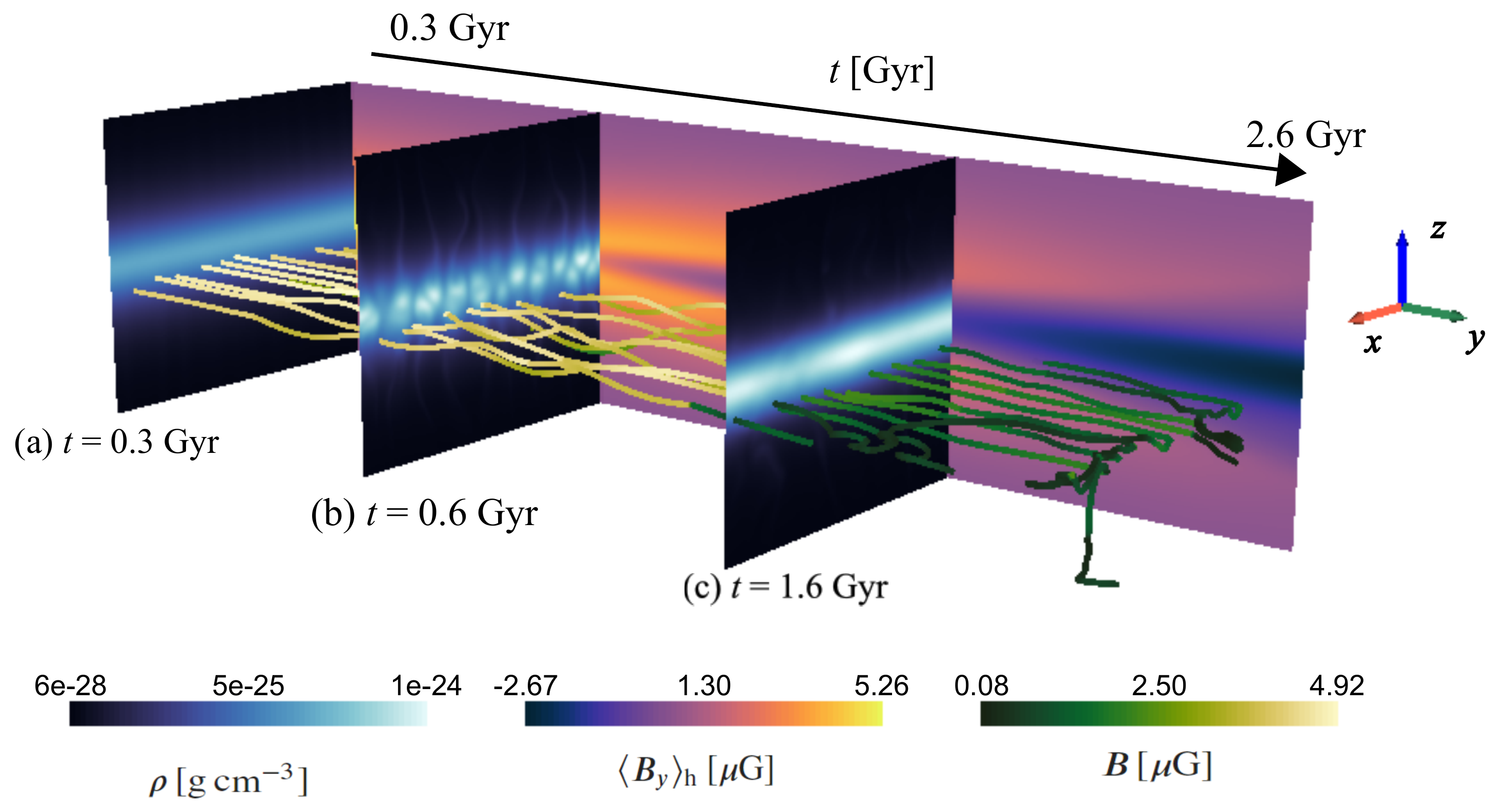}
    \caption{The evolution of the gas density and magnetic field in Model \SimB\ is illustrated for its three significant epochs: \textbf{(a)}~the linear stage, \textbf{(b)}~beginning of the magnetic field reversal in the early nonlinear stage and  \textbf{(c)}~the advanced nonlinear state (the specific simulation times are indicated for each frame). Selections of magnetic lines are shown (with colour representing the local magnetic field strength in $\upmu$G) in the $(x,y,z)$-space at the time indicated to the left of each frame. The horizontal average of the azimuthal magnetic field $\meanh{B_y}$ in $\upmu$G is shown with colour on the vertical $(z,t)$-plane as it evolves continuously (rather than at discrete times used for the magnetic lines). The gas density distribution is shown with colour on the vertical $(x,z)$-planes (in $\!\g\cm^{-3}$) for each time.}
    \label{fig:3d_vi_simb}
\end{figure*}
%-----------------------------------------

The presence of rotation only affects the momentum and induction equations,
so equations (1), (4)--(6), (9) and (10)
for the mass conservation and cosmic rays
of \citet{SPI} still apply and only the momentum and induction equations are augmented with terms containing $\Omega$ and $S$:
\begin{align}
\dderiv{\vec{u}}{t}& = -\frac{\nabla  P}{\rho} + \vec{g}+ \frac{(\nabla\times\vec{B})\times\vec{B}}{4\pi\rho} -S u_x \hat{\vec{y}} - 2\vec{\Omega} \times \vec {u} + \nabla\cdot\vec{\tau}\,,\label{N-S}\\ 
\deriv{\vec{a}}{t}&=\vec{u}\times(\nabla\times\vec{A})- Sa_y \hat{\vec x} - S x \deriv{\vec a}{y} - \eta\nabla\times(\nabla\times\vec{a})\label{ind}  \,,
\end{align}
where $\text{D}/\text{D}t= \partial/\partial t+(\vec{U}_0+\vec{u})\cdot\nabla$ is the Lagrangian derivative,
$\vec{g}$ is the gravitational acceleration and $\vec{\tau}$ is the viscous stress tensor. 
The Kepler gauge for the vector potential, as described by \citet{Oishi2011} \citep[see also][]{Brandenburg1995}, is appropriate for this shearing box framework.

We use the gravity field $\vec{g}=-g(z)\hat{\vec{z}}$ obtained by \citet{kg89} for the Solar vicinity of the Milky Way and consider an isothermal gas with the sound speed $c\sound= 18\kms$ and temperature $T = 3.2\times10^4\K$. In the background state (identified with the subscript zero, this is also the initial state), both the magnetic and cosmic ray pressures are adopted to be half the thermal pressure, $P_\text{m,0}/P_\text{th,0}= P_\text{cr,0}/P_\text{th,0}=0.5$, where $P_\text{th,0} = c\sound^2 \rho_0(0)$, $P_\text{m,0} = B_0^2(0)/(8 \pi)$ and $P_\text{cr,0} = \epsilon_\text{cr0}(0)/3$ are the thermal, magnetic and cosmic ray pressures, respectively, and $B_0(0)=5\muG$. The gas viscosity $\nu$ (included in $\vec{\tau}$) and magnetic diffusivity $\eta$ are chosen as $\nu = 0.1\kpc\kms$ and $\eta = 0.03\kpc\kms$, respectively, to be somewhat smaller than the turbulent values in the ISM \citep[see][for further details and justification]{SPI}. 

Table~\ref{tab:sims} presents the simulation runs discussed in this paper. The value of $\Omega$ near the Sun is close to $30\kms\kpc^{-1}$ (referred to as the nominal value hereafter), and $S=-\Omega$ when the rotational speed is independent of the galactocentric distance (a flat rotation curve), $|\vec{\Omega}\times\vec{r}|=\const$.
Model~\SimA\ is identical to Model~Sim6 of \citet{SPI}, Model~\SimD\ only differs by the solid-body rotation at the nominal angular velocity, Model~\SimB\ adds the large-scale velocity shear (differential rotation), whereas Model~\SimC\ has both the angular velocity and its shear doubled.
The averages at $z=\const$ (horizontal averages) are denoted $\meanh{\cdots}$. 

Figure~\ref{fig:3d_vi_simb} presents a pictorial summary of the changes in the magnetic field and gas density as the instability develops through its linear stage and then saturates in Model~\SimB. During the linear phase, at $t=0.3\Gyr$, the magnetic field and gas density retain the structure of the imposed fields with weak perturbations in $\rho$. By the weakly nonlinear stage at $t=0.6\Gyr$, both the gas density and magnetic field are strongly perturbed to the extent that the mean azimuthal magnetic field $\meanh{B_y}$ starts reversing. The reversal is complete in the late nonlinear stage at $t=1.6\Gyr$ and magnetic loops are prominent. We explain and detail these processes below.

%------------------------------------------------------------------------
\begin{figure}
    \centering
    \includegraphics[width=0.9\columnwidth]{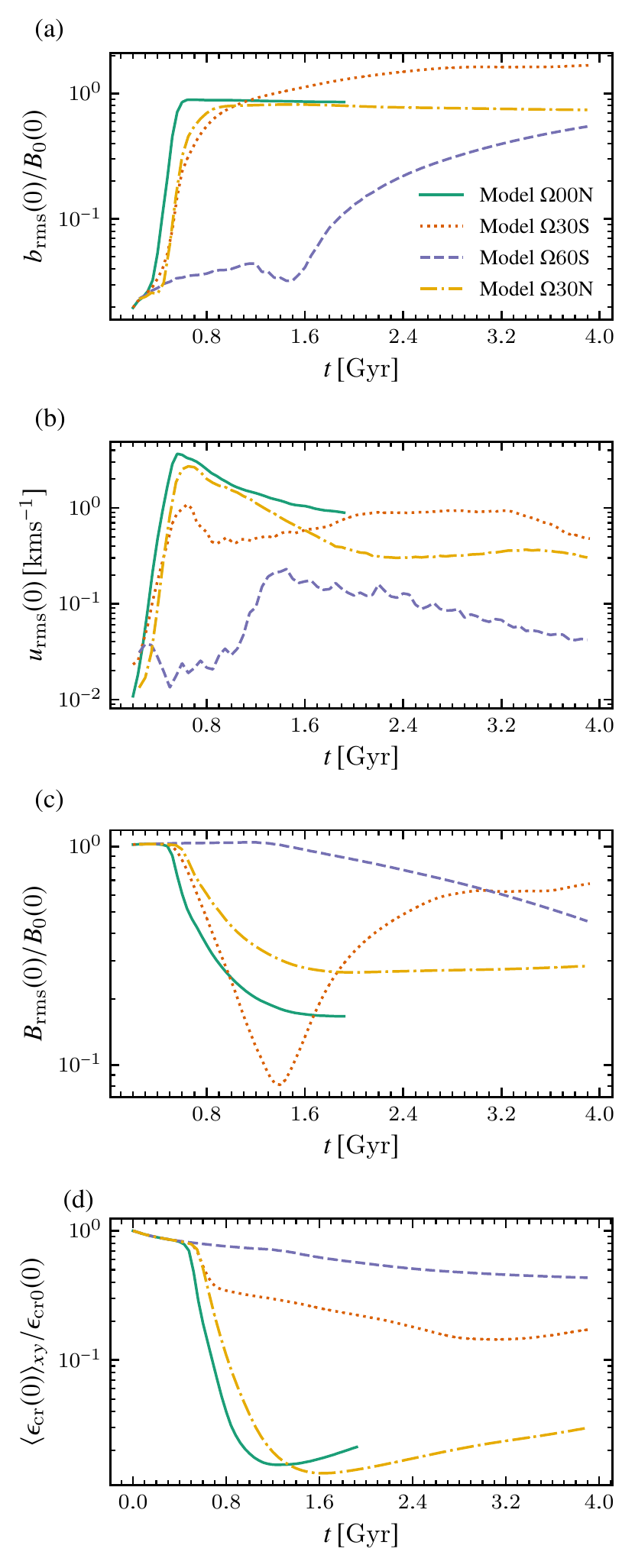}
    \caption{The evolution of the root-mean-square magnitudes
at the midplane $z=0$ of \textbf{(a)}~the magnetic field perturbation $|\vec{b}|$, normalised to $B_0(0)$ (the strength of the background magnetic field at $z=0$), and \textbf{(b)}~gas speed in the Models \SimA\ (solid, no rotation), \SimD\ (dash-dotted, solid-body rotation at the nominal $\Omega$), \SimB\ (dotted, differential rotation at the nominal $\Omega$ and $S$) and \SimC\ (dashed, doubled $\Omega$ and $S$).  Similarly, panels~\textbf{(c)} and \textbf{(d)} show the horizontally averaged total magnetic and cosmic ray energy densities at $z=0$ for those models, normalized to the respective midplane values in the background state, $\langle B\rangle_{xy}(0)/B_0(0)$ and $\langle\ecr\rangle_{xy}(0)/\ecri(0)$, respectively.
}
    \label{fig:gr_rot}
\end{figure}
%---------------------------------------------------------------------

%-----------------------------------------------------------------------
%\section{Results}\label{sec:re_rot}
\section{The linear instability}\label{sec:LI_rot}
The linear phase of the Parker instability in the absence of rotation is discussed in detail in our previous work \citep{SPI}, where we compare the growth rate and the spatial structure of the most rapidly growing mode with those obtained in a range of analytical and numerical models. In this section, we focus on the modifications of the exponentially growing perturbations caused by the rotation and velocity shear.

Figures~\ref{fig:gr_rot}a,b show the evolution (in both the linear and nonlinear stages) of the root-mean-square (\rms) magnitudes of the perturbations in the magnetic field and velocity, while Panels~(c) and (d) show how the total magnetic field strength $B_\rms$ and the mean cosmic ray energy density $\ecr$ at $z=0$, respectively, evolve in the models of Table~\ref{tab:sims}. As expected \citep{Shu1974, ZwKu1975, FogTag1994, FogTag1995, Hanasz1997a}, the instability growth rate $\Gamma$ (given in Table~\ref{tab:sims}) decreases systematically with the angular velocity. The stretching of the magnetic lines along the radial ($x$) direction by the Coriolis force enhances the magnetic tension thus opposing the instability while the differential rotation shears the perturbations to reduce the radial wavelength also suppressing the instability \citep{FogTag1994}. 

%---------------------------------------------------------------
\begin{figure}
    \centering
\includegraphics[width=0.995\columnwidth]{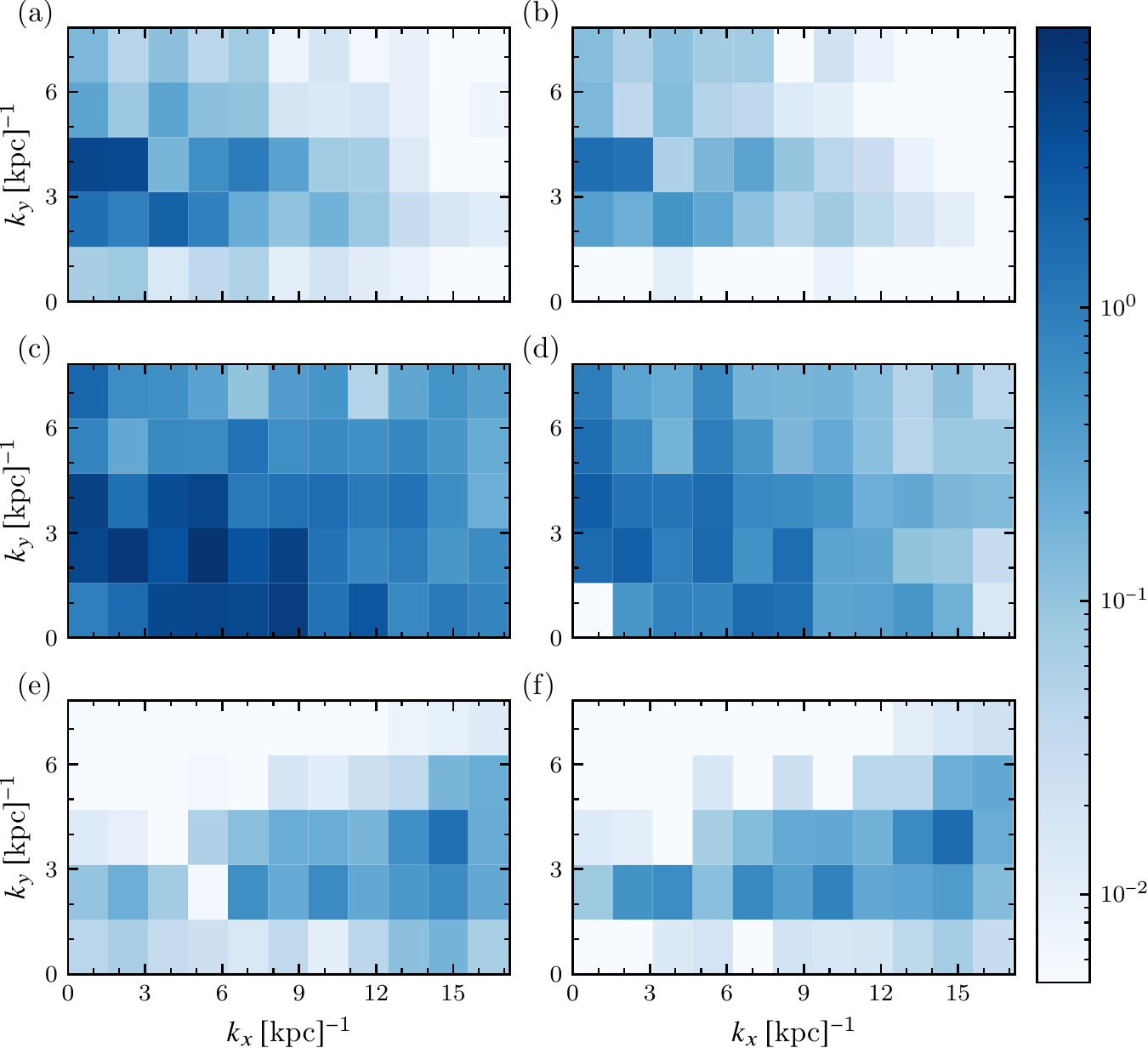}
    \caption{The two-dimensional power spectra of $u_z$ (left column, in the units of kpc$^2\,$km$^{2}\,$s$^{-2}$) and $b_z$ (right column, in kpc$^2\,\upmu$G$^2$), averaged over $|z|<1.75\kpc$, in Models~\SimA\ (a--b), \SimD\ (c--d) and \SimB\ (e--f) at $t=0.3\Gyr$ (the linear stage of the instability).}
    \label{fig:2dpsd}
\end{figure}
%---------------------------------------------------------

The spatial structure of the unstable modes is illustrated in Fig.~\ref{fig:2dpsd}, which presents the two-dimensional power spectra of the perturbations affected by the solid-body (c--d) and differential (e--f) rotation and compares them with the non-rotating case (a--b). The spectra of the velocity and magnetic field perturbations are identical when $\Omega=0$ but noticeable differences develop in rotating systems. In agreement with the analysis of \citet{Shu1974}, the dominant azimuthal wave number $k_y$ decreases under the influence of rotation. The solid-body rotation leads to wider spectra in the radial and azimuthal wave numbers, consistent with the weaker variation of the instability growth rate with $k_y$ in a rotating system  \citep[Fig.~1 of][]{FogTag1994}. Since the Coriolis force couples the radial and azimuthal motions, the spectra in $k_x$ and $k_y$ are more similar to each other than in the case $\Omega=0$. However, the velocity shear strongly reduces the range of $k_y$ while the perturbations have significantly larger radial wave numbers $k_x$ than in the cases $\Omega=0$ and $S=0$. 

%-------------------------------------------------------------------
\section{The saturated state}\label{sec:NI_rot}
Figure~\ref{fig:gr_rot} also shows that the nonlinear development of the instability and its statistically steady state are strongly affected by the rotation and velocity shear. Solid-body rotation does not affect much the magnitude of the magnetic field perturbations at $t\gtrsim 1\Gyr$, presented with the solid and dash-dotted curves in Panel~(a), but reduces the velocity perturbations shown in Panel~(b). Understandably, the velocity shear enhances both (the dotted curves) by stretching the radial magnetic fields which, in turn, affect the motions. The case of faster rotation and correspondingly stronger shear confirms this tendency (dashed curves). 

Panels (c) and (d) of Fig.~\ref{fig:gr_rot}, which show the total magnetic field strength and cosmic ray energy density at $z=0$, suggest that the structure of the magnetic field is changed profoundly by rotation and, especially, by the velocity shear. 
For example, the magnitude of the magnetic field perturbations in Model~\SimB\ shown with the dotted curve in Panel~(a) is less than twice larger than at $\Omega=0$ (solid curve), but the total magnetic field at $z=0$ shown in Panel~(c) is almost an order of magnitude stronger since the perturbation is better localised near $z=0$ (see below).
The instability still removes both the magnetic field and cosmic rays from the system as in the case $\Omega=0$, but at a much lower efficiency that depends on both the angular velocity and the rotational shear.

% -----------------------------------------------------------------------
\begin{figure*}
    \centering
\includegraphics[width=0.9\textwidth]{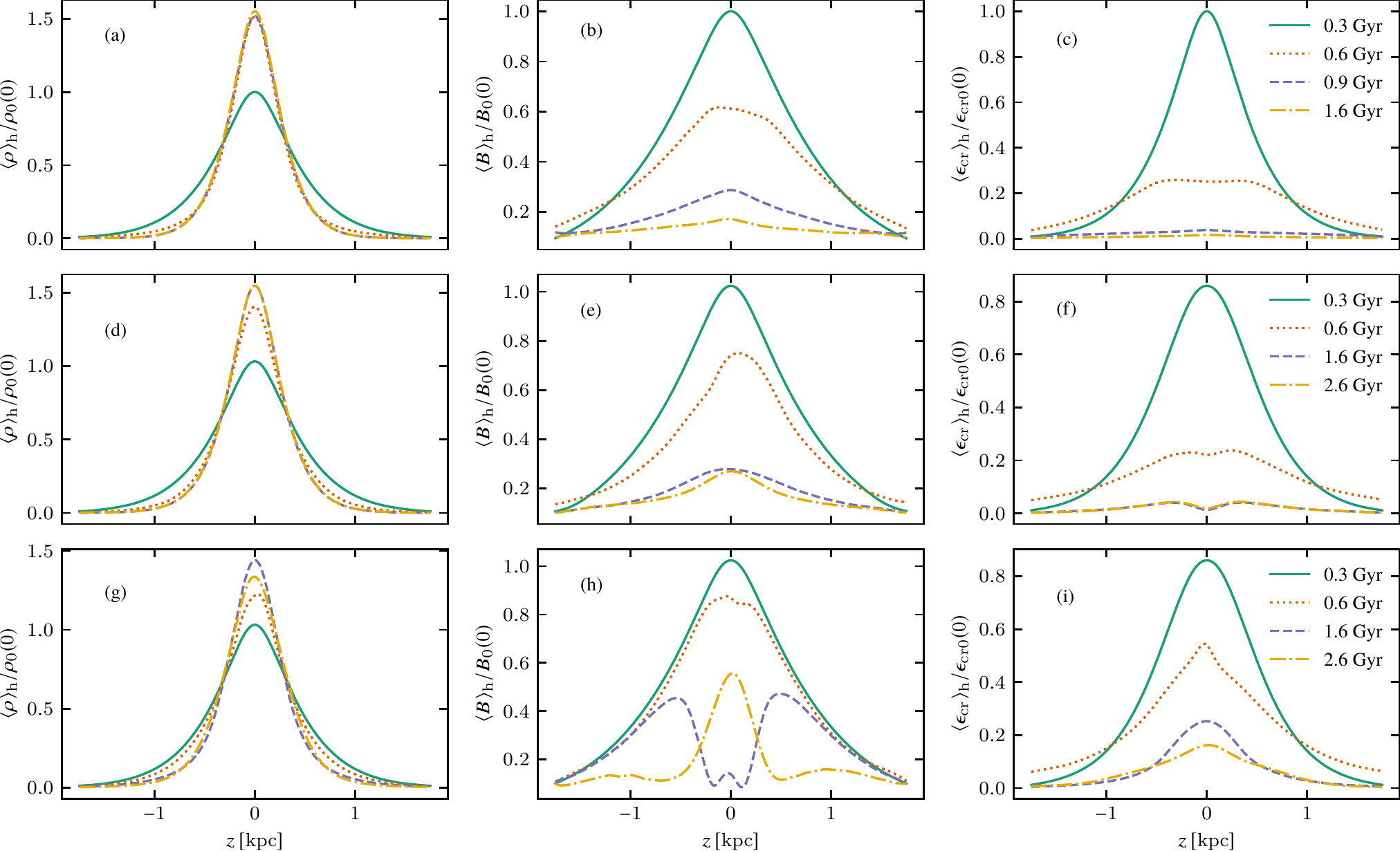}
    \caption{The evolution of the vertical profiles of the horizontally averaged and normalised gas density  $\meanh{\rho}/\rho_0(0)$ (left-hand column), magnetic field strength $\meanh{B}/B_0(0)$ (middle) and cosmic ray energy density $\meanh{\ecr}/\ecri(0)$ (right-hand column). First row: Model \SimA\ (no rotation), second row: Model~\SimD\ (nominal solid-body rotation), third row: Model~\SimB\ (nominal rotation and shear). The times corresponding to the line styles are given in the legend of each row.
    Note that the direction of the mean azimuthal magnetic field $\meanh{B_y}$ has reversed within a certain distance of the midplane at the later times, $t=1.6$ and $2.6\Gyr$.}
    \label{fig:profs}
\end{figure*}
% ------------------------------------------------------

%-------------------------------------------------------------
\begin{figure*}
    \centering
\includegraphics[width=0.7\textwidth]{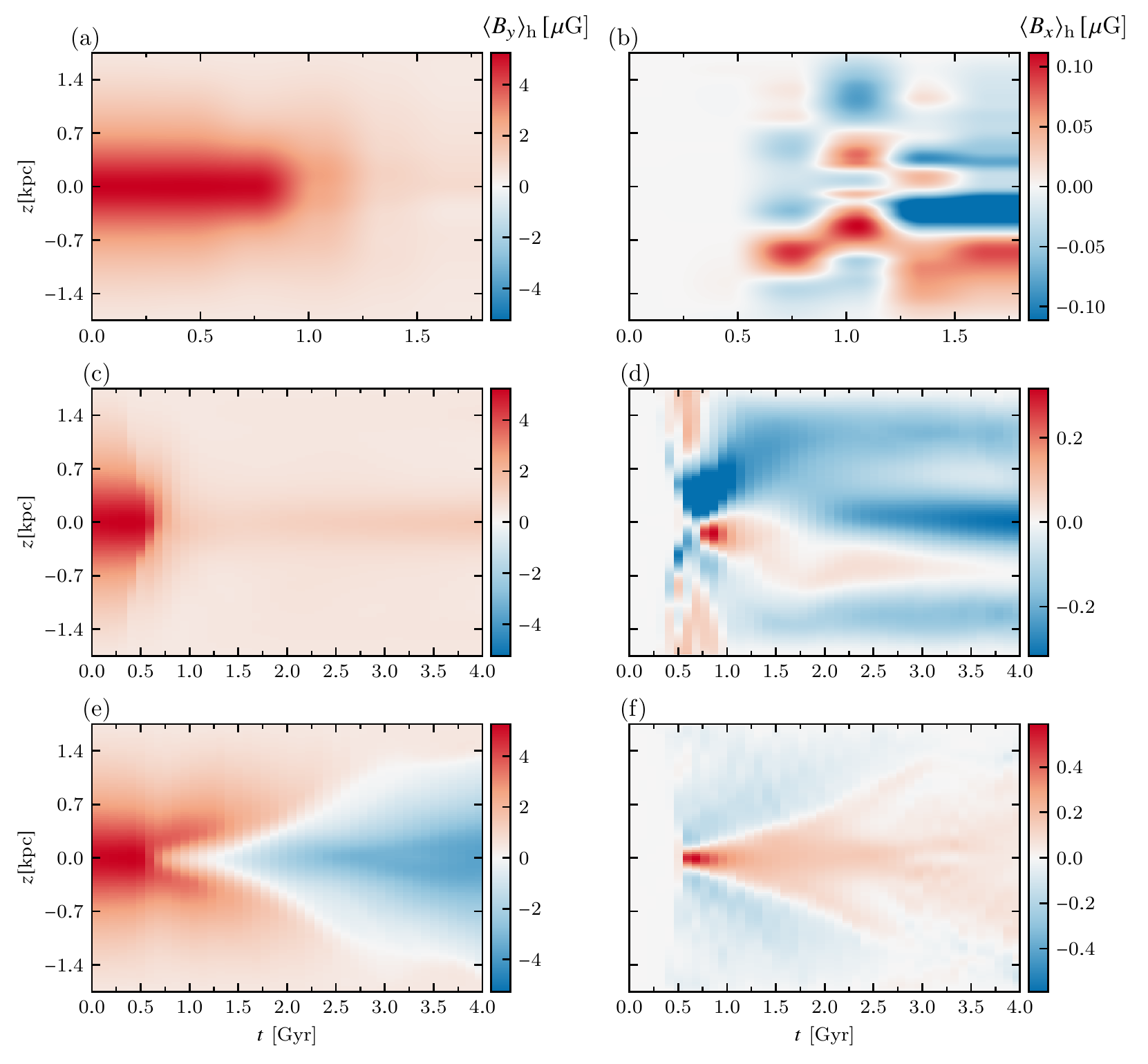}
    \caption{The evolution of the horizontally averaged magnetic field components, $\meanh{B_y}$ (left-hand column) and $\meanh{B_x}$ (right-hand column) in Models  \SimA\ (a--b), \SimD\ (c--d) and \SimB\ (e--f).
    For \SimB\ the mean azimuthal field $\meanh{B_y}$ decreases after $t=0.6\Gyr$, and undergoes a reversal in sign at $t\approx 1.6\Gyr$, 
    with the reversal then spreading to higher altitudes. 
    Meanwhile, the mean radial field $\meanh{B_x}$ becomes positive and relatively strong near $z=0$ rather abruptly at $t\approx0.5\Gyr$ and then also spreads away from the midplane.}
    \label{fig:bxbyxy}
\end{figure*}
%------------------------------------------------------------

As compared to the case $\Omega=0$, the system retains stronger magnetic field under the solid-body rotation but less cosmic rays, as shown with the solid and dash-dotted curves in Fig.~\ref{fig:gr_rot}(c,d). Figure~\ref{fig:profs} clarifies the details of the changes effected by rotation and velocity shear, presenting the varying vertical profiles of the gas density, magnetic fields and cosmic rays in Models \SimA, \SimD\ and \SimB. Both solid-body and differential rotations reduce the gas scale height in the saturated state. The comparison of Panels (b--c) and (e--f) shows that the solid-body rotation leads to narrower distributions (smaller scale heights) of both magnetic field and cosmic rays about the midplane. Moreover, as we discuss below, the gas flow becomes helical in a rotating system (see Section~\ref{sec:tc}), supporting the mean-field dynamo action. As a result, a large-scale radial magnetic field $B_x$, clearly visible in Fig.~\ref{fig:bxbyxy}(d,f), emerges in a rotating system.

The velocity shear changes the nonlinear state qualitatively. Firstly, the scale heights of $B$ and $\ecr$ near the midplane are even smaller at $t=0.6\text{--}0.9\Gyr$ in Panels (h) and (i) than at the comparable times in Panels (e) and (f). Secondly, and more importantly, the vertical profile of the magnetic field strength evolves to become more complicated at $t=1.6\Gyr$ in Panel (h), and the cosmic ray distribution reflects this change. The energy density of cosmic rays in Model~\SimB, $\meanh{\ecr}(0)=0.2\ecri$ at $t=1.6\Gyr$ (Fig.~\ref{fig:profs}i) is ten time larger than in Model~\SimA. Differential rotation helps to confine cosmic rays because it drives dynamo action generating strong horizontal magnetic field, and this slows down the escape of cosmic rays as they spread along larger distances guided by the magnetic field.

The change in the vertical profile of $\meanh{B}$ in Model~\SimB\ at $t=1.6\Gyr$ reflects the reversal of the horizontal magnetic field near the midplane discussed and explained in Section~\ref{sec:mr}.

% ------------------------------------------------------
\section{Magnetic field reversal}\label{sec:mr}
The reversal of the magnetic field in the nonlinear stage of the instability has been noticed earlier by a few authors (see Section~\ref{In}) but our simulations identify it as a generic feature 
of the Parker and magnetic buoyancy instabilities in rotating systems. 
This process is illustrated in Fig.~\ref{fig:bxbyxy} which shows how the evolution of the large-scale horizontal magnetic field components $\meanh{B_x}$ and $\meanh{B_y}$ depends on rotation and the velocity shear.

Figure~\ref{fig:bxbyxy}a shows again \citep[see also][for details]{SPI} that,  in a non-rotating system, the azimuthal magnetic field $\meanh{B_y}$ decreases with time in strength and its scale height increases, while the radial field $\meanh{B_x}$ shown in Fig.~\ref{fig:bxbyxy}b is much weaker and varies along $z$ without any systematic pattern. Solid-body rotation causes two major changes: the azimuthal field strength (Fig.~\ref{fig:bxbyxy}c) first decreases faster than without rotation but then starts growing and, at late times, is stronger than for $\Omega=0$. The field direction remains the same as of the imposed field, $\meanh{B_y}>0$. Meanwhile, the radial field (Fig.~\ref{fig:bxbyxy}d) is, at late times, comparable in strength to $\meanh{B_y}$, well-ordered and is predominantly negative, $\meanh{B_x}<0$. This change is a result of the mean-field $\alpha^2$-dynamo action driven by the mean helicity of the gas flow as discussed in Section~\ref{sec:tc}.

The differential rotation of Model~\SimB\ (Fig.~\ref{fig:bxbyxy}e,f) changes the evolution even more dramatically: it drives the more efficient $\alpha\omega$-dynamo with stronger $\meanh{B_x}$ and, remarkably, exhibits a reversal of the large-scale horizontal magnetic field. The reversal starts in the weakly nonlinear phase at $t=0.5\Gyr$ with a rather abrupt emergence of a relatively strong positive radial magnetic field near the midplane, $\meanh{B_x}>0$. The velocity shear with $S<0$ stretches the positive radial field into a negative azimuthal magnetic field, so that $\meanh{B_y}$ starts decreasing and reverses at $t=1.6\Gyr$ (Fig.~\ref{fig:bxbyxy}e). 
The total horizontal magnetic field strength $(\meanh{B_x}^2+\meanh{B_y}^2)^{1/2}$ decreases to a minimum before increasing again, as $\meanh{B_y}$ decreases to zero and then re-emerges with the opposite direction.
These changes in the large-scale magnetic field structure start near the midplane and spread to larger altitudes
because of the magnetic buoyancy.

% %-------------------------------------------------------
\begin{figure}
    \centering
    \includegraphics[width=0.8\columnwidth]{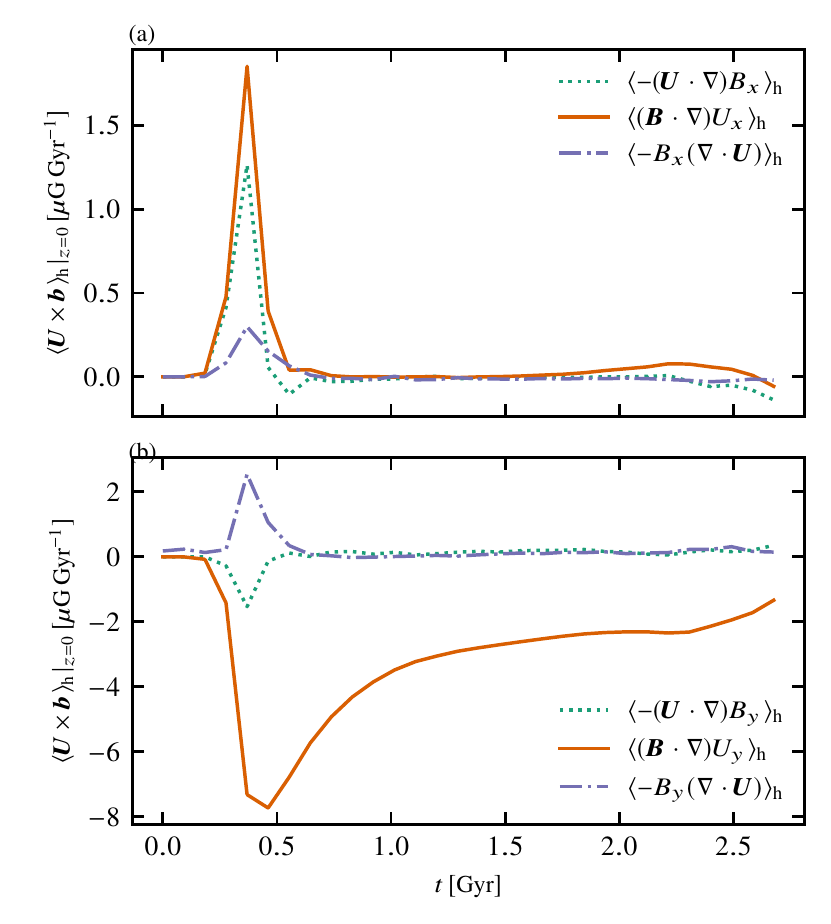}
    \caption{The evolution of the three terms on the right-hand side of the induction equation \eqref{indB} volume-averaged near the midplane ($z<0.4\kpc$): \textbf{(a)}~the radial ($x$) and \textbf{(b)}~the azimuthal ($y$) components of the stretching term  $(\vec B \cdot \vec \nabla)\vec{U}$ (solid), advection $-(\vec U \cdot \vec \nabla)\vec{B}$ (dotted) and compression $-B (\vec \nabla \cdot \vec U)$ (dash-dotted), in model \SimB.}
    \label{fig:ind}
\end{figure}
%---------------------------------------------------------
\begin{figure}
    \centering
    \includegraphics[width=0.8\columnwidth]{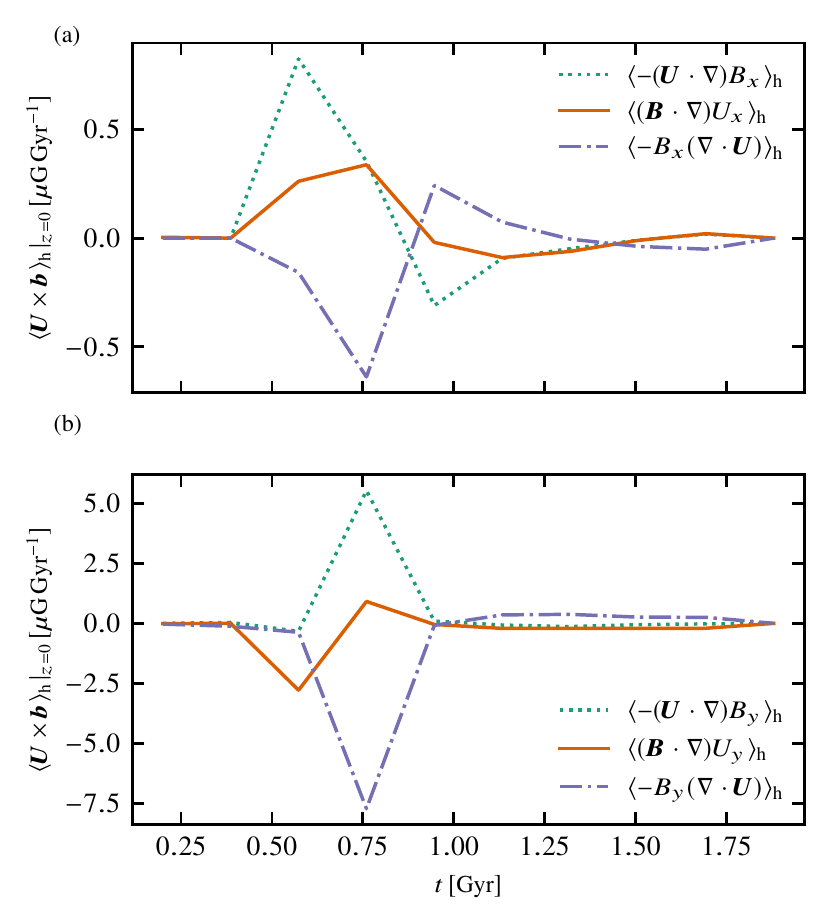}
    \caption{As in Fig.~\ref{fig:ind}, but for model \SimA.}
    \label{fig:ind_norot}
\end{figure}
%----------------------------------------------------------
\begin{figure}
    \centering
    \includegraphics[width=0.9\columnwidth]{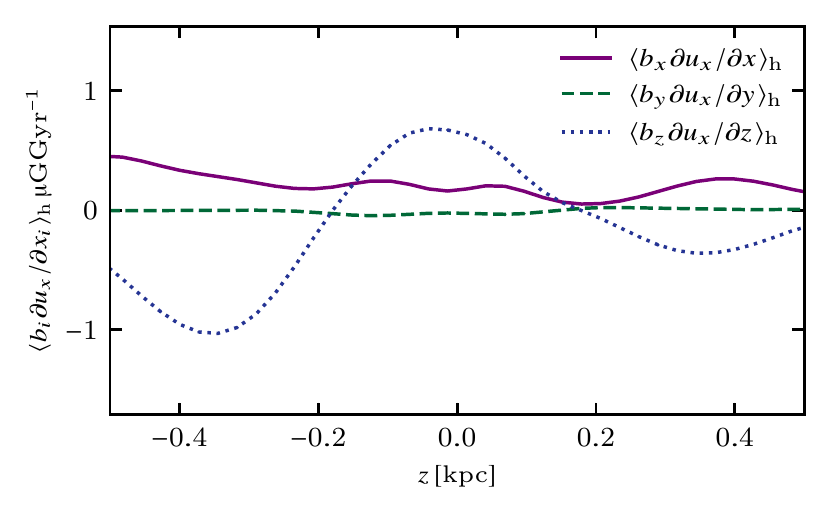}
    \caption{The vertical variation of the horizontally averaged stretching terms in equation~\eqref{str_x} in Model~\SimB\ at $t=0.7\Gyr$ near the midplane: $\meanh{b_x\,\partial u_x/\partial x}$ (solid), $\meanh{b_y\,\partial u_x/\partial y}$ (dashed) and $\meanh{b_z\,\partial u_x/\partial z}$ (dotted).}
    \label{fig:bxrhs}
\end{figure}
%---------------------------------------------------------

%----------------------------------------
\subsection{The mechanism of the reversal}\label{MR}
To understand the process that leads to the reversal of the large-scale azimuthal magnetic field, we
consider individual terms in the induction equation
written for the deviation from the imposed
magnetic field,

\begin{equation}\label{indB}
\deriv{\vec{b}}{t}=-(\vec{U}\cdot\nabla)\vec{B} +(\vec{B}\cdot\nabla)\vec{U} - \vec{B}\nabla\cdot\vec{U} + \eta\nabla^2 \vec{b}\,.
\end{equation}
Figure~\ref{fig:ind} shows, for Model~\SimB, the evolution of the mean radial and azimuthal components of the first three terms on the right-hand side of this equation, which represent the advection, stretching and compression of the corresponding magnetic field components near the midplane. 
The stretching terms $(\vec{B}\cdot\nabla)U_x$ and $(\vec{B}\cdot\nabla)U_y$ clearly dominate, producing a mean radial field $\meanh{B_x}>0$ during the weakly nonlinear stage, $0.6\lesssim t\lesssim 0.8\Gyr$, 
which decreases only slowly at later times (because of diffusion and buoyancy) while being gradually stretched by the differential rotation $S<0$ into a negative azimuthal field $\meanh{B_y}$, eventually leading to the reversal of the initially positive $\meanh{B_y}$.  
This picture is very different from that for Model~\SimA, where the stretching terms in both components rapidly vanish after a negative excursion during the early nonlinear phase (see Figs~\ref{fig:bxbyxy}a,b and \ref{fig:ind_norot}). 
Under the solid-body rotation, a positive radial field does emerge near $z=0$ in the early nonlinear stage but, without the velocity shear, this does not lead to the reversal of the azimuthal field (Fig.~\ref{fig:bxbyxy}c,d).

We have analyzed various parts of the averaged stretching term $\meanh{(\vec{B}\cdot\nabla)U_x}$ in the $x$-component of equation~\eqref{indB} to understand which of them produces a positive radial component of the mean field.
We note that $\meanh{U_x}=0$ and then $\meanh{(\vec{B}\cdot\nabla)U_x}=\meanh{(\vec{b}\cdot\nabla)u_x}$.
Thus,
\begin{equation}\label{str_x}
\meanh{(\vec{B}\cdot\nabla)U_x}=\Meanh{b_x\deriv{u_x}{x}} +\Meanh{b_y\deriv{u_x}{y}} +\Meanh{b_z\deriv{u_x}{z}}\,.
\end{equation}
Figure~\ref{fig:bxrhs} shows that the first two terms on the right-hand side of this equation are less significant than the third term, and that $\meanh{b_z\,\partial{u_x}/\partial{z}}>0$ at $|z|\lesssim0.2\kpc$. 
The term $\meanh{b_x\,\partial{u_x}/\partial{x}}$ also contributes to the generation of a positive $\meanh{B_x}$ at all $z$.

%-----------------------------------------------
\begin{figure}
    \centering
    \includegraphics[width=0.8\columnwidth]{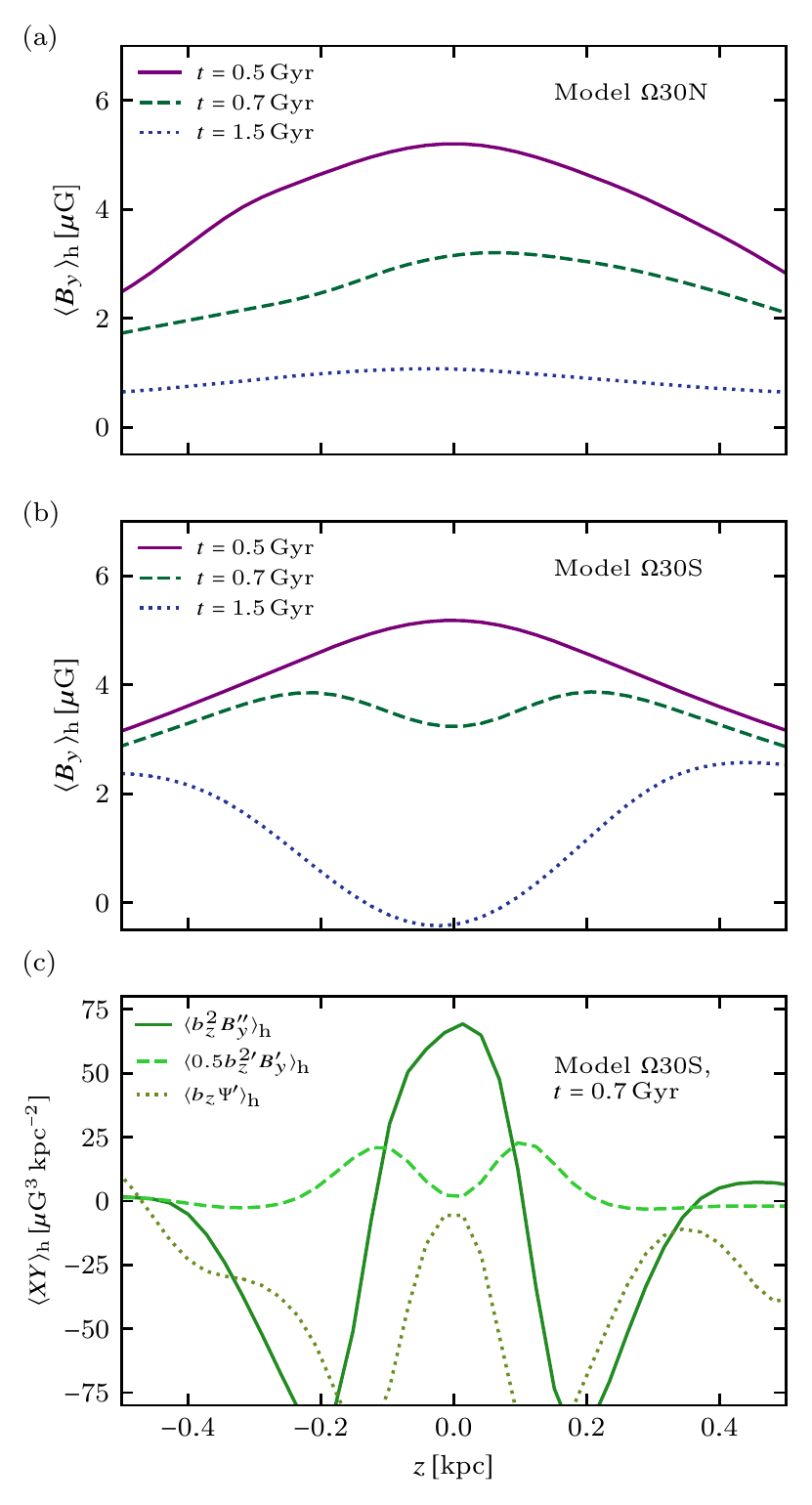}
        \caption{The vertical profiles of the horizontally averaged azimuthal field, $\meanh{B_y}$, at $=0.5\Gyr$ (solid), $t=0.7\Gyr$ (dashed)  and $1.5\Gyr$ (dotted) in Models \textbf{(a)}~\SimD\ and \textbf{(b)}~\SimB. Panel \textbf{(c)} shows the variation with $z$ of the correlations on the right-hand-side of equation~\eqref{eq:duxdz} for Model~\SimB\  at $t=0.7\, \Gyr$: $\meanh{(b_z)^2\,\partial^2 B_y/\partial z^2}$ (solid), $\meanh{\tfrac12\partial(b_z)^2/\partial z\,\partial B_y/\partial z}$  (dashed) and $\meanh{b_z\, \partial\Psi/\partial z}$ (dotted).}
    \label{fig:vpby}
\end{figure}
%-----------------------------------------------------

The positive correlation between $b_z$ and $\partial u_x/\partial z$, the main driver in the generation of the positive $\meanh{B_x}$, arises because of: (i)~the Coriolis force; and (ii)~the emergence of a local minimum of $\meanh{B_y}$ at the midplane produced by the buoyancy.
To demonstrate this, we express $u_x$ using the $y$-component of the  momentum equation~\eqref{N-S} with $S=-\Omega$, differentiate the result with respect to $z$, 
multiply it by $b_z$ and average to obtain
%----------------------------------
\begin{align}\label{eq:duxdz}
\rho\Omega\Meanh{b_z\deriv{u_x}{z}} 
&=\frac{1}{4\pi}\Meanh{b_z^2\deriv{^2B_y}{z^2}} +\frac{1}{8\pi}\Meanh{\deriv{b_z^2}{z}\deriv{B_y}{z}}
\nonumber\\
&+\Meanh{b_z\,\deriv{\Psi}{z}-b_z\,\deriv{\rho}{z}\Omega u_x}\,,
\end{align}
%-----------
%-------------------------------
where we have neglected the fluctuations in $\rho$ when averaging on the left-hand side (which is justifiable since the random gas speed is subsonic) and $\Psi$ combines all other terms:
%---------------------
\begin{equation}\label{Psi}
\Psi=-\rho\dderiv{u_y}{t} -\deriv{P}{y} -\frac{1}{8\pi}\deriv{b^2}{y}+\frac{1}{4\pi}\left(b_x\deriv{b_y}{x} +b_y\deriv{b_y}{y}\right)\,,
\end{equation}
 %-------------------------------
where we neglect the viscosity (represented by the viscous stress tensor $\vec{\tau}$) and $b^2=b_x^2+b_y^2+b_z^2$. Figures~\ref{fig:vpby}a,b show vertical profiles of $\meanh{B_y}$ in Models~\SimD\ (where no reversal occurs) and \SimB, while Fig.~\ref{fig:vpby}c clarifies the form of various terms in equation~\eqref{eq:duxdz}. The positive correlation $\meanh{b_z\,\partial u_x/\partial z}$ emerges because of the first term on the right-hand side as soon as magnetic buoyancy produces a local minimum of $\meanh{B_y}$ at $z=0$ (see Fig.~\ref{fig:vpby}b), so that $\partial^2 B_y/\partial z^2$ is systematically positive at $z=0$. Such a minimum does not develop in the case of solid-body rotation (Fig.~\ref{fig:vpby}a) where no reversal of $\meanh{B_y}$ happens.  As shown in Fig.~\ref{fig:vpby}c, the second and third terms in  equation~\eqref{eq:duxdz} are smaller in magnitude than the first term near $z=0$ and partially compensate each other.
The correlation $\meanh{b_z\,\partial u_x/\partial z}$ is dominant and positive near $z=0$, driving a reversal in the large-scale magnetic field near the midplane which then spreads to larger $|z|$ as shown in Fig.~\ref{fig:ind}e,f because of the magnetic buoyancy.  We stress that the minimum of $\meanh{B_y}$ at $z=0$ can only arise at the nonlinear stage of the instability, because only then do the fluctuations $b_y$ not average to zero.

We have verified that the reversal is not sensitive to the direction of the imposed magnetic field  $B_0(z)\hat{\vec{y}}$; i.e., it occurs in the exactly the same manner for $B_0(z)>0$ and $B_0(z)<0$.
Our simulations extend to $4\Gyr$ in duration (see Fig.~\ref{fig:bxbyxy}). 
This is already a significant fraction of the galactic lifetime; therefore, we did not extend them further to find out if a further reversals would occur at later times. 
However, periodic reversals occur in a similar model where the unstable magnetic field is generated by an imposed \textit{mean-field dynamo action} (Y.~Qazi et al.\ 2022, in preparation). 
It appears that the emergence of the local minimum of $\meanh{B_y}$ at $z=0$ and its ensuing reversal is related to the mean-field dynamo action
(which our imposed field emulates). 
The dynamo is driven by the mean helicity of the gas flow, and both Models~\SimD\ and \SimB\ support
this mechanism (as discussed below). 
However, the dynamo in Model~\SimD, which has solid-body rotation (so is an $\alpha^2$-dynamo), is too weak, whereas the differential rotation of Model~\SimB\ enhances the dynamo enough (making it an $\alpha\omega$-dynamo) to produce the reversal.
In the next section, we compute and discuss the mean helicity of the gas flow and other evidence for the mean-field dynamo action in Model~\SimB.

%------------------------------------------------------------
\section{Helicity and dynamo action}\label{sec:tc}
In Models~\SimD, \SimB\  and \SimC, the Coriolis force causes the gas motions to become helical, and the resulting $\alpha$-effect produces a large-scale radial magnetic field $\meanh{B_x}$ \citep[e.g., Sect.~7.1 of][]{ShSu22}. 
Differential rotation (in Models~\SimB\ and \SimC) 
enhances the dynamo significantly, and we have discovered that this leads to a reversal in the azimuthal magnetic field direction discussed in Section~\ref{sec:mr}. Both types of the turbulent dynamo ($\alpha^2$ dynamo in \SimD\ and $\alpha\omega$ in \SimB\ and \SimC) are driven by the mean kinetic helicity of the gas flow $\chi\kin=\mean{\vec{\uh}\cdot(\nabla\times\vec{\uh})}$, and the current helicity of the magnetic fluctuations $\chi\m=\mean{\vec{\bh}\cdot(\nabla\times\vec{\bh})}$ opposes the dynamo instability leading to a reduction of the $\alpha$-coefficient until a steady state is achieved \citep[e.g., Sect.~7.11 of][]{ShSu22}. Here overbar denotes a suitable averaging, and we use the horizontal averages in our discussion, 
so $\vec{\uh}$ and $\vec{\bh}$ are understood as the deviations from the horizontal averages 
$\meanh{\vec{B}}$ and $\meanh{\vec{U}}$, 
such that 
\begin{equation}\label{buh}
\vec{B}=\meanh{\vec{B}}+\vec{\bh}\,, \quad \vec{U}=\meanh{\vec{U}}+\vec{\uh}\,, \quad \Meanh{\vec{\bh}}=0\,, \quad  \Meanh{\vec{\uh}}=0\,.
\end{equation}

%--------------------------------------------------------
\begin{figure}
    \centering
    \includegraphics[width=\columnwidth]{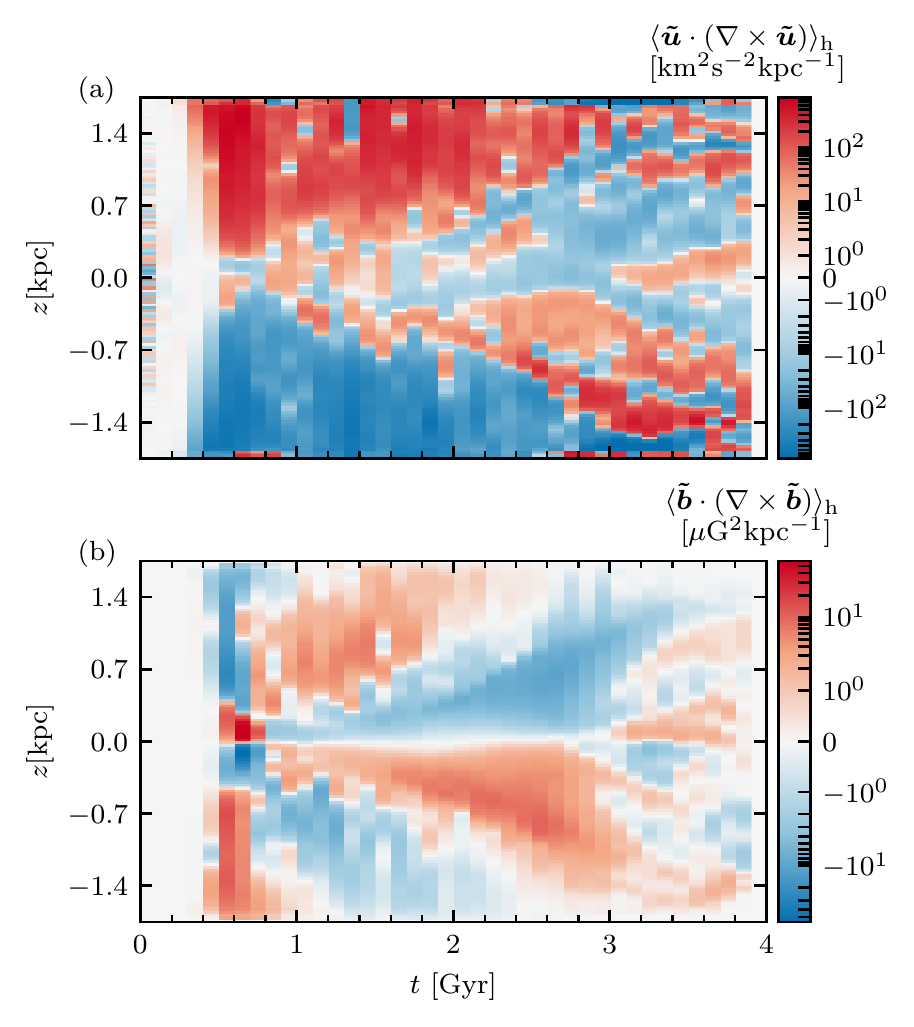}
    \caption{The evolution of the horizontally averaged \textbf{(a)}~kinetic helicity $\meanh{\vec{\uh}\cdot(\nabla \times \vec{\uh})}$
    and \textbf{(b)}~current helicity $\meanh{\vec{\bh}\cdot(\nabla\times\vec{\bh}})$ in Model~\SimB.}
    \label{fig:helt}
\end{figure}
%-------------------------------------------------------------------

%------------------------------------------------------------------------------
\begin{figure}
    \centering
    \includegraphics[width=0.85\columnwidth]{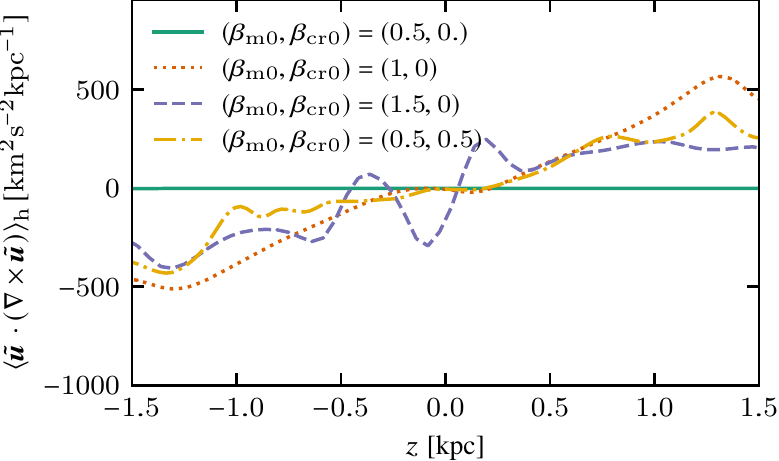}
    \caption{The spatial distribution of the mean kinetic helicity $\chi\kin$ at $t=0.7\Gyr$ for four imposed (initial) magnetic field strengths specified by the parameters $\beta_\text{m,0}$ and $\beta_\text{cr,0}$ defined in equation~\eqref{PmPc} and given in the legend. Among the models shown in this figure, cosmic rays are present only in Model~\SimB\ where $(\beta_\text{m,0},\beta_\text{cr,0})=(0.5, 0.5)$ (dash-dotted: this is a vertical cross-section of the distribution in Fig.~\ref{fig:helt}a).}
    \label{fig:alphab0}
\label{alphakz}
\end{figure}
%--------------------------------------------------------------------------

Figure~\ref{fig:helt} shows the evolution of the kinetic and current helicities and their variation with $z$ obtained using the horizontal averages. As expected, both quantities have odd symmetry in $z$ \citep[e.g., Sect.~11.3.1 of][]{ShSu22}. 
Both are weak throughout the linear phase when the instability-driven perturbations are still weak, but increase significantly in magnitude during the early nonlinear phase at about $t=0.5\Gyr$. The kinetic helicity reaches its maximum magnitude $|\chi\kin|=|\meanh{\vec{\uh}\cdot(\nabla\times\vec{\uh}}|= 851\km^2\s^{-2}\kpc^{-1}$ near the upper and lower boundaries,  $z=\pm1.6\kpc$, during the transitional phase at $t=0.6\Gyr$. At a later time, $t=1.9\Gyr$, the kinetic helicity reduces to a maximum of $|\chi\kin|= 340\km^2\s^{-2}\kpc^{-1}$  at $|z|=1.6\kpc$. 
At early stages of the evolution, the current helicity has local extrema close to the midplane, where the magnetic field is stronger, $|\chi\m|=|\meanh{\vec{\bh}\cdot(\nabla\times\vec{\bh})}|=89\muG^2\kpc^{-1}$ at $t=0.6\Gyr$, $|z|=0.1\kpc$. The extrema move away from the midplane in the nonlinear stage, to reach $|\chi\m|=7\muG^2\kpc^{-1}$ at $t=1.2\Gyr$, $|z|=0.5\kpc$ and $|\chi\m|=5\muG^2\kpc^{-1}$ at $t=3\Gyr$, $|z|=1\kpc$.

The vertical profiles of both kinetic and current helicities evolve in a rather complicated manner, 
with $\chi\kin<0$ at $z>0$ close to the midplane 
(although the magnitude is small),
and $\chi\kin>0$ at larger $z$  in the case of pure magnetic buoyancy (dotted curve in Fig.~\ref{alphakz} representing $t=0.7\Gyr$). In Model~\SimB, $\chi\kin<0$ at $z>0$ close to the midplane just before $t=0.7\Gyr$. Negative $\chi\kin$ at $z>0$ is expected from the action of the Coriolis force on the ascending and descending volume elements \citep[Sect.~7.1 of][]{ShSu22}. However, $\chi\kin>0$, as it occurs at larger $z$ for all models presented in Fig.~\ref{alphakz}, is unexpected (see below for a discussion).

%--------------------------------------------------------
\begin{figure}
    \centering
    \includegraphics[width=0.8\columnwidth]{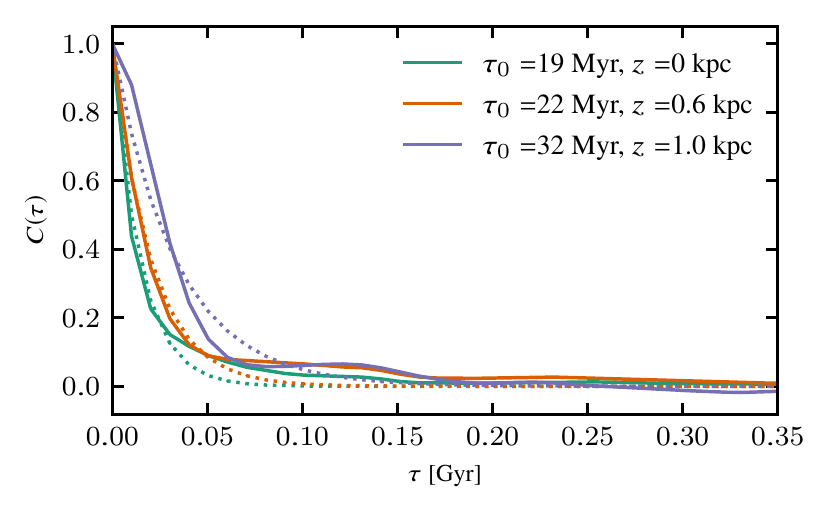}
    \caption{The time autocorrelation function of the vertical velocity component, equation~\eqref{Ctau}, for $0\leq t\leq 2\Gyr$ (with the minimum time lag of $10\Myr$) at $z=0$ (solid), $0.6$ (dashed) and $1\kpc$ (dotted) in Model~\SimB. The correlation time $\tau_0$ at each $z$ is given in the legend, 
    obtained from the fits of the form  $C(\tau)=\exp(-\tau/\tau_0)$, shown with dotted curves.
}
    \label{fig:uztcorr}
\end{figure}
%-----------------------------------------------------------------
 
%--------------------------------------------
\begin{figure}
    \centering
    \includegraphics[width=\columnwidth]{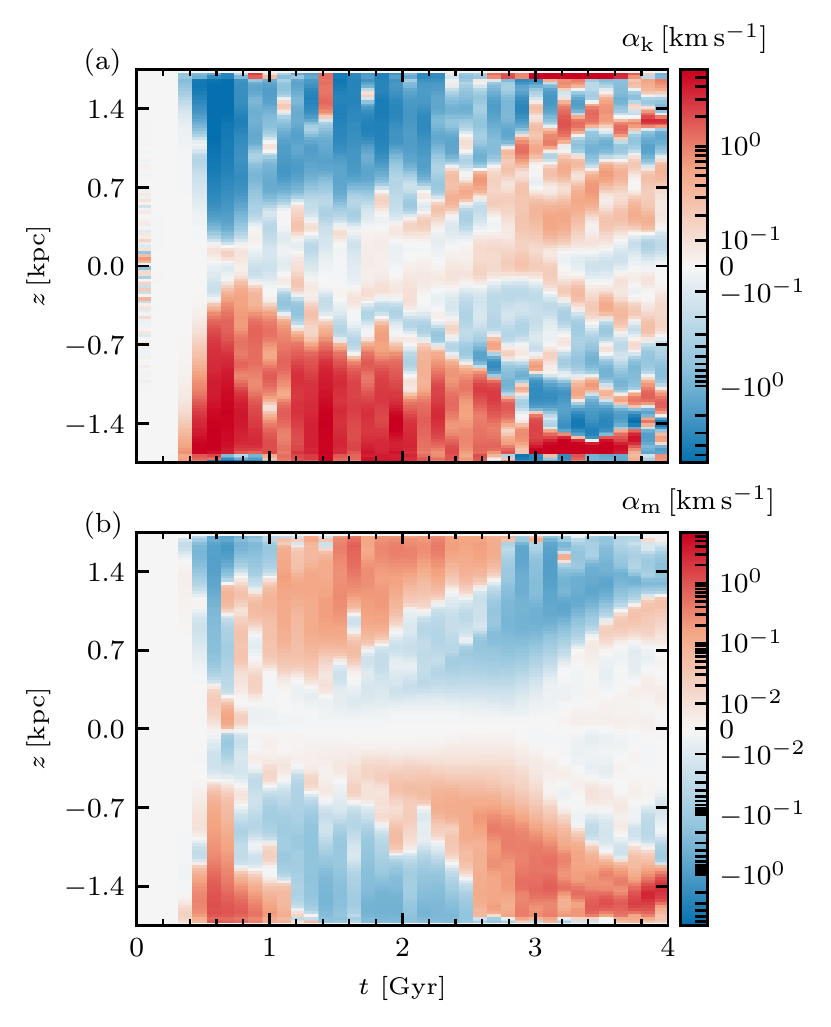}
    \caption{The evolution of \textbf{(a)}~$\alpha\kin$ and \textbf{(b)}~$\alpha\m$, given in equations~\eqref{akm}, in Model~\SimB.}
    \label{fig:alpha}
\end{figure}
%-----------------------------------------------------------------------

The $\alpha$-coefficient of the nonlinear mean-field dynamo is related to the kinetic and current helicities as  \citep[Sect.~7.11.2 of][]{ShSu22}
\begin{equation}\label{alpha}
\alpha=\alpha\kin+\alpha\m\,,
\end{equation}
where, in terms of the horizontal averages,
\begin{equation}\label{akm}
\alpha\kin=-\tfrac13\tau_0\meanh{\vec{\uh}\cdot(\nabla\times\vec{\uh})}\,,
\qquad
\alpha\m=\tfrac13\tau_0\frac{\meanh{\vec{\bh}\cdot(\nabla\times\vec{\bh})}}{4\pi\rho}\,,
\end{equation}
and $\tau_0$ is the characteristic (correlation) time of the random flow. 

The relevant time scale $\tau_0$ differs from the time scale of the linear instability $2\pi/(u_0 k_y)$ where $u_0$ and $k_y$ are the characteristic speed and azimuthal wave number of the most unstable mode shown in Figs~\ref{fig:gr_rot}b and Fig.~\ref{fig:2dpsd}e--f, respectively. Instead, $\tau_0$ is determined by nonlinear effects and has to be measured separately. We calculate the correlation time using the time autocorrelation function $C(\tau)$ of $u_z$ (the vertical velocity $u_z$ is a representative component since it is directly related to the instability),
\begin{equation}\label{tcorr}
\tau_0=\int_0^\infty C(\tau)\,\dd\tau\,,  
\end{equation}
with the normalized autocorrelation function calculated as
\begin{equation}\label{Ctau}
C(\tau)=\frac{1}{T\Meanh{\uh_z^2}}\Meanh{\int_0^T \uh_z(t,\vec{x}) \uh_z(t+\tau,\vec{x})\,\dd t}\,,
\end{equation}
where $T$ is the duration of the time series used to compute $C(\tau)$.
For a given $z$, the integral in equation~\eqref{Ctau} is calculated for each $(x,y)$ and the result is averaged over $(x,y)$.
Thus defined, the autocorrelation function and the corresponding correlation time depend on $z$.

Figure~\ref{fig:uztcorr} shows the time autocorrelation of $u_z$ at three values of $z$, 
and the form $C(\tau)=\exp(-\tau/\tau_0)$ provides a good fit, with the fitted values of $\tau_0$ given in the legend: they vary between $18\Myr$ at $z=0$ and $40\Myr$ at $z=1.5\kpc$. We use the fitted $C(\tau)$ to estimate $\tau_0$ as this provides a more accurate result than the direct integration as in the definition~\eqref{tcorr}.

We use $\tau_0=30\Myr$ in equations~\eqref{akm}, and the results are shown in Fig.~\ref{fig:alpha}.
The largest in magnitude values $|\alpha\kin|\approx 7\kms$ are reached during the transition phase around $t=0.6\Gyr$ near $|z|=1.5\kpc$, whereas $|\alpha\m|$ is at its maximum around $3\kms$ during the nonlinear phase at $t=3.6\Gyr$. 

%-----------------------------------------------------------
\begin{figure}
    \centering
    \includegraphics[width=0.85\columnwidth]{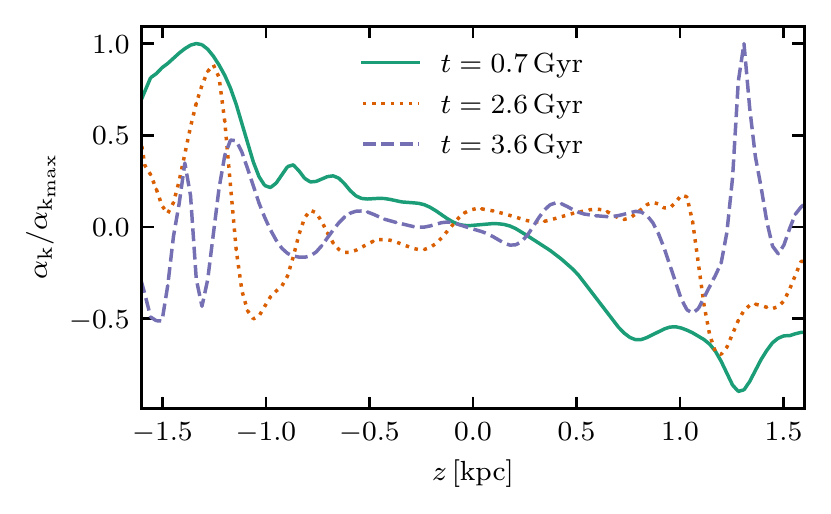}
    \caption{The variation of the normalised $\alpha\kin$ with $z$ in the early ($t=0.7\Gyr$, solid) and late ($t=2.6\Gyr$, dotted, $t=3.6\Gyr$, dashed) nonlinear stages in Model~\SimB.}
    \label{alphak_2t}
\end{figure}
%--------------------------------------------------------------

The spatial structure of $\alpha\kin$ is relatively simple during the early nonlinear phase but becomes more complicated later. 
Closer to the midplane and at later stages of the evolution,
$\alpha\kin>0$ at $z>0$ (and $\alpha\kin<0$ at $z<0$) as expected, and the region where $\alpha\kin$ is predominantly positive 
(albeit small in magnitude)
extends to larger $|z|$ with time (see Fig.~\ref{alphak_2t} representing vertical sections of Fig.~\ref{fig:alpha}a).

As expected, the sign of the current helicity is opposite to that of $\alpha\kin$ at almost all $z$ and $t$, 
so that the back-reaction of the magnetic field on the flow weakens the dynamo action leading to a (statistically) steady state at $t\gtrsim3\Gyr$.

The negative sign of $\alpha\kin$ at $z>0$ (corresponding to the positive
kinetic helicity $\chi\kin$) appears to be a specific feature of a system
driven by magnetic buoyancy or another magnetically driven instability such as
the magneto-rotational instability (MRI). \citet{Hanasz1998} argue, using a
model of reconnecting magnetic flux ropes, that negative $\alpha\kin$ at $z>0$
can occur in magnetic buoyancy-driven mean-field dynamos. In his analysis of
the mean electromotive force produced by the magnetic buoyancy instability in
its linear stage, \citet[][his Fig.~4]{The00a} finds $\alpha<0$ in the
unstable region of the northern hemisphere in spherical geometry (corresponding
to $z>0$ in our case), although the `anomalous' sign of $\alpha\kin$ remained
unnoticed \citep{The00b}. However, \citet{BrSc98} find $\alpha\kin>0$ at $z>0$
in their analysis of the $\alpha$-effect due to magnetic buoyancy.
\citet{BrSo02} find  $\alpha\kin<0$ at $z>0$ in simulations of the MRI-driven
dynamos (their Section~2 and $\alpha_{yy}$ in Figs~5, 7, 9 and 11). Kinetic
helicity (and the corresponding $\alpha\kin$) of this `anomalous' sign is also
found in the simulations of MRI-driven dynamos of P.~Dhang et al.\ (2023, in
preparation) (K.~Subramanian 2022, private communication). The origin and
properties of the kinetic helicity of random flows driven by magnetic buoyancy
and MRI deserves further attention. Our results indicate not only that the
kinetic helicity has the anomalous sign but also that it can change in space
and time.

%--------------------------------------------
\begin{figure}
    \centering
    \includegraphics[width=\columnwidth]{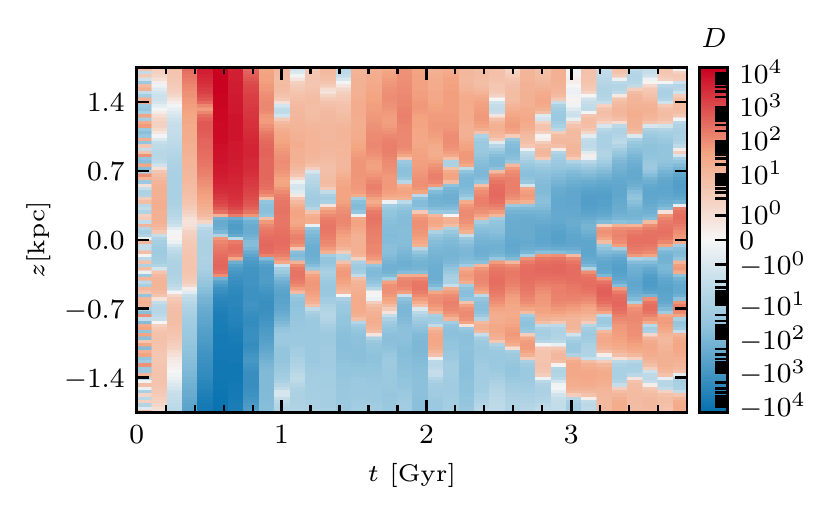}
    \caption{The evolution and vertical variation of the dynamo number of equation~\eqref{DN} in Model~\SimB.}
    \label{DynNum}
\end{figure}
%------------------------------------------------

The current helicity (Fig.~\ref{fig:helt}b) and the corresponding contribution to the $\alpha$-effect (Fig.~\ref{fig:alpha}b) have the opposite signs to, and closely follow both the spatial distribution and evolution of, $\chi\kin$ and $\alpha\kin$ respectively (although the magnetic quantities have smoother spatial distributions than the corresponding kinetic ones). This confirms that the action of the Lorentz force on the flow weakens the dynamo action as expressed by equation~\eqref{alpha}. Together with the removal of the large-scale magnetic field by the Parker instability, this leads to the eventual evolution of the system to the statistically steady state.

Although the gas flows that become helical are driven by the instability, no simple and obvious relation of the mean helicity to the parameters that control the strength of the instability is apparent. Figure~\ref{alphakz} shows how the vertical profile of the kinetic helicity $\chi\kin$ changes with the magnetic and cosmic ray pressures in the initial (imposed) state, specified in terms of their ratios to the thermal pressure at $z=0$,
\begin{equation}\label{PmPc}
 \beta_\text{m0} = \frac{B_0(0)^2}{8\pi c\sound^2\rho_0(0)} 
 \quad\text{and}\quad 
 \beta_\text{cr0} = \frac{(\gamma\cra-1)\ecri(0)}{c\sound^2\rho_0(0)}\,,
\end{equation}
where $\gamma\cra=4/3$. 
To avoid complications associated with the cosmic rays in the system behaviour, only one model of the four illustrated in Fig.~\ref{alphakz} contains cosmic rays (Model~\SimB\ discussed elsewhere in the text). The midplane strengths of the imposed magnetic field $B_0(0)$ corresponding to $\beta_\text{m0}=0.5,1$ and 1.5 are 5, 7 and $9\muG$, respectively.  When $(\beta_\text{m0},\beta_\text{cr0})=(0.5,0)$, the magnetic field is too weak to be unstable and the system remains in the state of magneto-hydrostatic equilibrium, $\chi\kin=0$. Adding cosmic rays,  $(\beta_\text{m0},\beta_\text{cr0})=(0.5,0.5)$ (Model~\SimB) destabilises the system producing helical flows discussed above. Adding magnetic rather than cosmic ray pressure, $(\beta_\text{m0},\beta_\text{cr0})=(1,0)$, also makes the system unstable, and the resulting mean helicity   at larger $|z|$ is greater than for $(\beta_\text{m0},\beta_\text{cr0})=(0.5,0.5)$. 
A still stronger magnetic field, $(\beta_\text{m0},\beta_\text{cr0})=(1.5,0)$ leads to comparable $\chi\kin$ the previous two cases in $|z|\lesssim1\kpc$, except near the midplane.
Altogether, it is difficult to identify a clear pattern in the dependence of the magnitude and spatial distribution of the mean helicity of the gas flow driven by the Parker instability; this invites further analysis, both analytical and numerical.

The dimensionless measure of the mean-field dynamo activity in a differentially rotating gas layer is provided by the dynamo number \citep[Section~11.2 of][]{ShSu22}
\begin{equation}\label{DN}
D = \frac{\alpha S h^3}{\beta^2}\,,
\end{equation}
where $h$ is the layer scale height, $S$ is the velocity shear rate ($S=-\Omega$ in our case), $\alpha$ is given in equation~\eqref{alpha} and 
\begin{equation}\label{beta}
\beta=\tfrac13\tau_0\meanh{\uh^2}+\eta
\end{equation}
is the magnetic diffusivity. The first term in this expression is the turbulent diffusivity and $\eta$ is the explicit magnetic diffusivity from equation~\eqref{ind} or \eqref{indB}. As we use the horizontal averages in these relations, $D$ is a function of $z$ and varies with time together with $h$, $\alpha$ and 
$\beta$;
thus defined, $D$ might be better called the local dynamo number, a measure of the dynamo efficiency at a given $z$ and $t$. In Model~\SimB, $\eta=0.03\kpc\kms$ while the turbulent diffusivity varies, at $t=1\Gyr$, from $0.03\kpc\kms$ at $z=0$ to $0.5\kpc\kms$ at $z=1\kpc$ (a nominal turbulent diffusivity in the ISM, where turbulence is mainly driven by supernovae, is $1\kpc\kms$). The dynamo amplifies a large-scale magnetic field provided $|D|>D\crit$, where $D\crit$ is a certain critical dynamo number (see below).

Figure~\ref{DynNum} shows how the dynamo number varies with $t$ and $z$. During the transient phase, $\meanh{\uh^2}$
is relatively low while $|\alpha|$ is at its maximum. The resulting dynamo number is as large as $|D|\simeq10^4$.  As the system evolves into the nonlinear state, the turbulent diffusivity increases and the dynamo number reduces in magnitude. At $t=0.6\Gyr$, $D$ varies from $4$ near the midplane to $6\times10^3$ at $z=1\kpc$. At later times, $D$ is larger near the midplane and reduces further in magnitude: at $t=0.9\Gyr$, $D=300$ near the midplane and 9 at  $z=1\kpc$.

As shown by \citet{RST80}, the $\alpha\omega$-dynamo in flat geometry generates oscillatory magnetic fields for $D>0$, quadrupolar for $D\gtrsim 180$ and dipolar for $D\gtrsim 550$. The behaviour of the large-scale  magnetic field in Model~\SimB\ is consistent with these results: it is quadrupolar and oscillatory.

% -------------------------------------------------------------
\begin{table}
\caption{The cross-correlation coefficient $r$ of the fluctuations in various energy densities in the statistically steady state of Model~\SimB\ at  $t=2.6\Gyr$ presented as $a,b$, where $a$ and $b$ refer to $z=0.5$ and $1\kpc$, respectively.}
\centering
\begin{tabular}{ccccc}
\hline
             &$\epsilon'\therm$  &$\epsilon'\cra$       &$\epsilon'\m$   &$\epsilon'\kin$\\
\hline
$\epsilon'\therm$   &$1,1$          &$0.2,-0.03$  &$-0.02,-0.2$  &$-0.14,0.12$\\
$\epsilon'\cra$         &               &$1,1$          &$-0.4,-0.8$   &$0.2,0.05$ \\
$\epsilon'\m$    &               &               &1,1            &$-0.29,-0.1$\\
$\epsilon'\kin$    &               &               &               &$1,1$   \\
\hline
    \end{tabular}
    \label{tab:corrcoef_full}
\end{table}
%---------------------------------------------------------------

%------------------------------------
\section{Relative distributions of cosmic rays and magnetic field}\label{RDCRMF}
Similar to our analysis in \citet{SPI}, we present in Table~\ref{tab:corrcoef_full} the Pearson cross-correlation coefficient between the fluctuations in energy densities for different components in model \SimB\ at $z=0.5$ and $1\kpc$ for the late nonlinear stage at $t=2.6\Gyr$, derived as
\begin{equation}\label{e_flucs}
\begin{split}
\epsilon'\m &= \frac{B^2-\Meanh{B^2}}{8\pi}\,, & \ecr'&= \ecr - \meanh{\ecr}\,,\\
\epsilon'\therm &= c\sound^2\left( \rho - \meanh{\rho}\right), & \epsilon\kin'&=\tfrac{1}{2}\rho \uh^2-\Meanh{\tfrac{1}{2}\rho \uh^2}\,.
\end{split}
\end{equation}
The only significant entry in the table is the anti-correlation between the magnetic and cosmic ray energy fluctuations at $z=1\kpc$ 
where their contribution to the total pressure is noticeable (see Section~\ref{VFFB}). 
There are no signs of energy equipartition between cosmic rays and magnetic fields at kiloparsec scales;
nor are there indications of equipartition at the turbulent scales, for either cosmic ray protons \citep{SSWBS18} or electrons \citep{TSSS22}.

%---------------------------------------------------------
\section{Vertical flows and force balance}\label{VFFB}

%---------------------------------------------------------
\begin{figure}
    \centering
    \includegraphics[width=0.9\columnwidth]{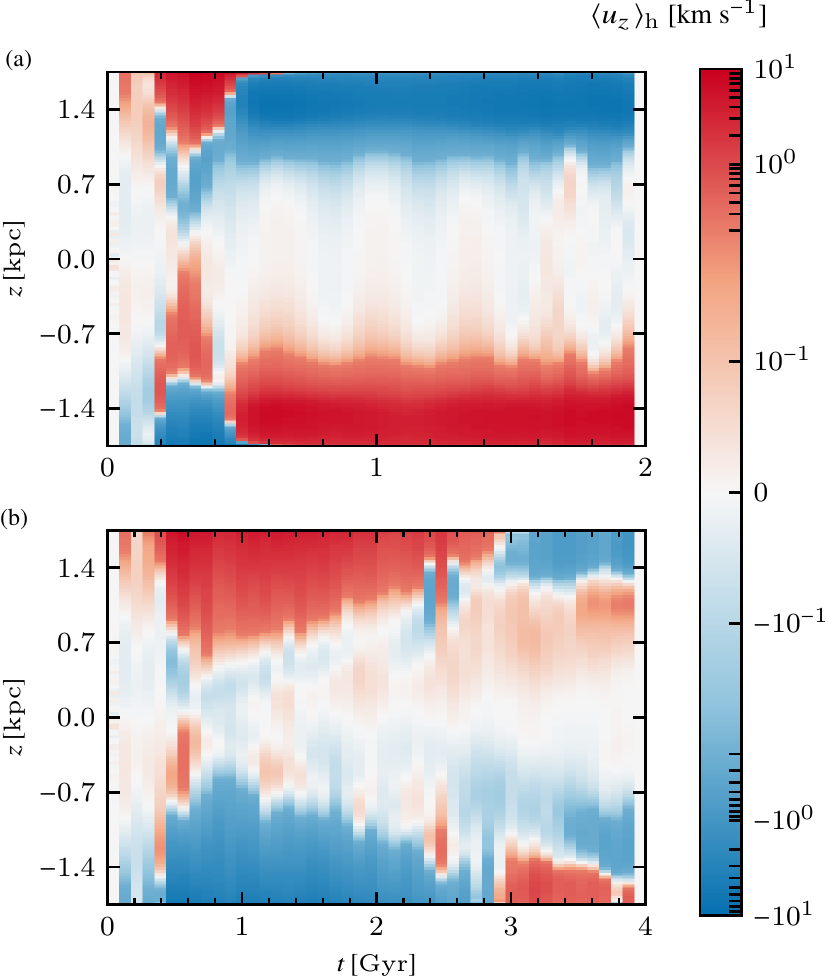}
    \caption{The evolution and variation with $z$ of the horizontally averaged vertical velocity $\meanh{u_z}$ in Models \textbf{(a)}~\SimD\ and \textbf{(b)}~\SimB.}
    \label{fig:kg89rotuz}
\end{figure}
%------------------------------------------

Rotation affects significantly the vertical gas flow driven by the instability. As discussed by \citet{SPI} (and also in Model~\SimA), a systematic gas outflow is transient without rotation and only occurs during the early nonlinear stage. Figure~\ref{fig:kg89rotuz} shows the horizontally averaged vertical velocity $\meanh{u_z}$ in Models~\SimD\ (solid-body rotation) and \SimB\ (differential rotation). In both cases, systematic vertical flows occur at $|z|\gtrsim1\kpc$. The solid-body rotation (Fig.~\ref{fig:kg89rotuz}a) does not change much the structure of the flow in comparison with the non-rotating system, with a transient outflow during the early nonlinear stage and a weak inflow at later times. 
In Model~\SimD, the maximum outflow speed is $|\meanh{u_z}|=9\kms$ at $t=0.7\Gyr$, followed by the inflow at the speed $|\meanh{u_z}|=7\kms$ at $t>1.4\Gyr$. However, differential rotation not only changes dramatically the magnetic field structure and evolution (Fig.~\ref{fig:bxbyxy}), but also supports a prolonged period of a systematic gas outflow at $0.6\lesssim t\lesssim3\Gyr$, which eventually evolves into a weak gas inflow at large $|z|$ (Fig.~\ref{fig:kg89rotuz}b). The maximum outflow speed in Model~\SimB\ is $|\meanh{u_z}|=7\kms$ at $t=0.6\Gyr$ at large $|z|$, while the later inflow speed is $|\meanh{u_z}|=1\kms$ at $t\gtrsim 3\Gyr$.
\label{vfb}

%--------------------------------------------------------
\begin{figure*}
    \centering
    \includegraphics[width=0.8\textwidth]{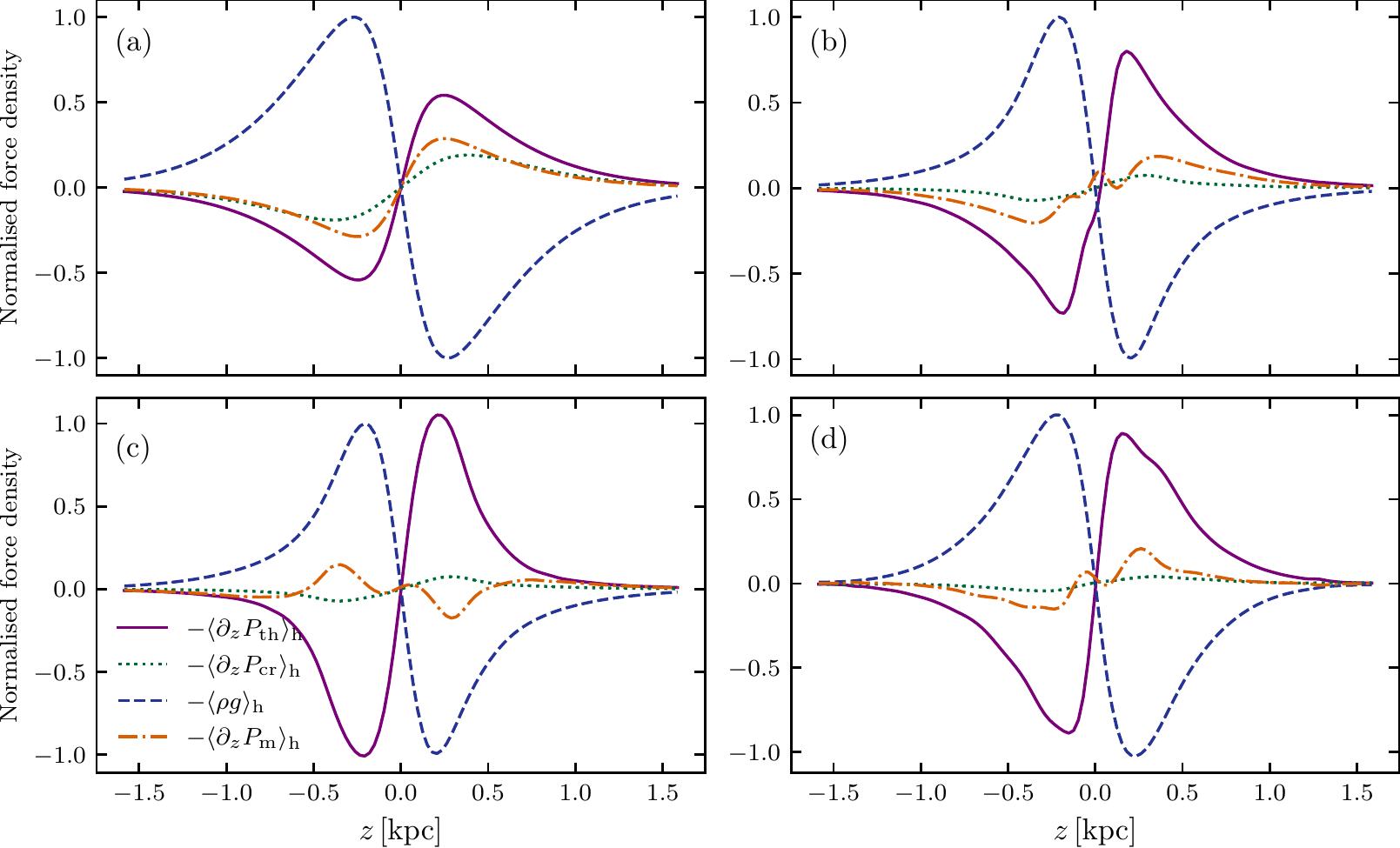}
    \caption{The vertical profiles of the horizontally averaged vertical forces in Model~\SimB\ normalised to the maximum magnitude of the gravitational force (dashed, repeated in all panels for reference): thermal (solid), cosmic ray (dotted) and magnetic (dash-dotted) pressure gradients. The contribution of the magnetic tension is much weaker, so it is not shown. Each panel represents a different evolutionary stage: \textbf{(a)}~$t=0.3\Gyr$ (linear instability), \textbf{(b)}~$0.6\Gyr$ (transitional);     \textbf{(c)}~$1.6\Gyr$ (nonlinear state when the magnetic field has just reversed near $z=0$) and  \textbf{(d)}~$3.6\Gyr$ (late nonlinear stage).
    }
    \label{fig:vfb}
\end{figure*}
%--------------------------------------------------------
% %---------------------------------------------
\begin{figure}
    \centering
    \includegraphics{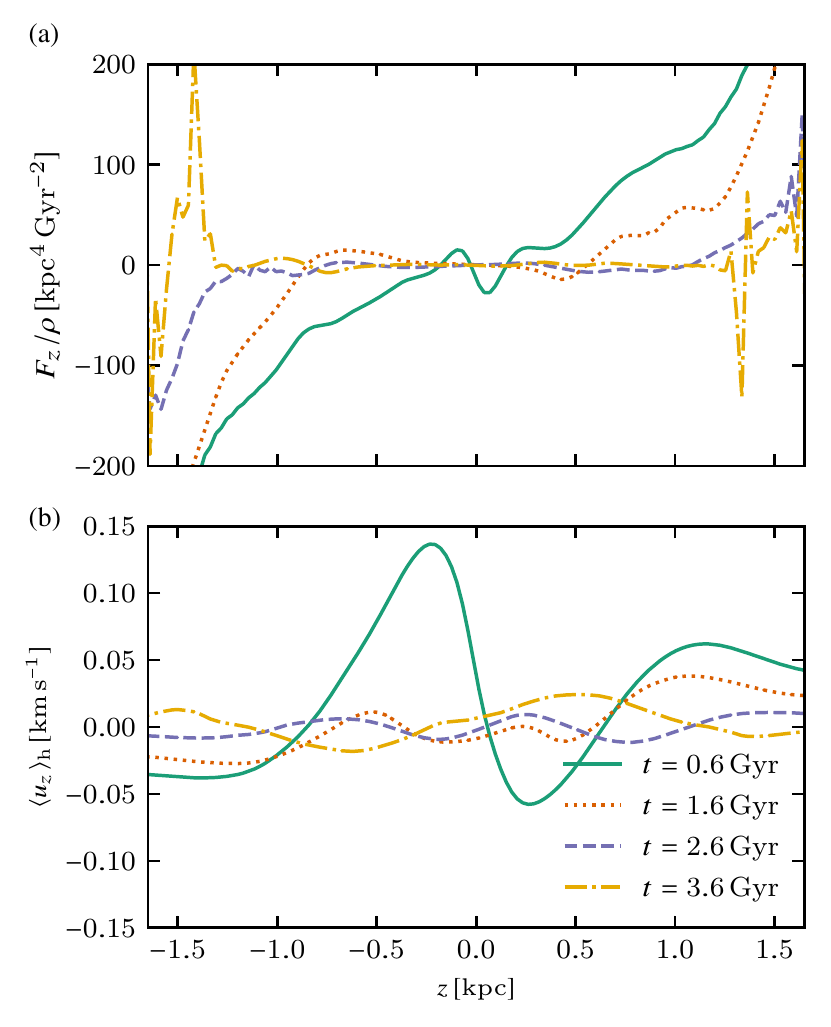}
    \caption{\textbf{(a)}~The total vertical force per unit mass and \textbf{(b)}~the resulting vertical velocity at times  $t=0.6$ (solid), $1.6$ (dotted), $2.6$ (dashed) and $3.6\Gyr$ (dash-dotted).
    }
    \label{fig:vf}
\end{figure}
% %-------------------------------------------------------

The pattern of the vertical flows shown in Fig.~\ref{fig:kg89rotuz}b is not dissimilar to the structure of the magnetic field shown in Fig.~\ref{fig:bxbyxy}e--f and the dynamo number (Fig.~\ref{DynNum}) 
--- especially at later stages, $t\gtrsim3\Gyr$ ---
suggesting that the magnetic field contributes noticeably to the vertical flow in Model~\SimB. 

To understand what drives the vertical flows, we present in Fig.~\ref{fig:vfb} the vertical forces acting during various evolutionary stages of Model~\SimB. It is instructive to compare them with those in non-rotating systems discussed by \citet{SPI}. Without rotation, as in Model~\SimA\ \citep[see also Fig.~12 of][]{SPI}, both magnetic and cosmic ray pressures are reduced significantly as the system evolves into the nonlinear state, and the vertical gas flows are driven by the thermal pressure gradient. 
This changes in Model~\SimB, where magnetic field, and to a lesser extent cosmic rays, make a stronger contribution to the force balance. 
Moreover, the gravity force and the thermal pressure gradient balance each other almost completely in the nonlinear state, so that the weaker magnetic and cosmic ray pressures appear to be capable of controlling the vertical velocity pattern, especially at $|z|\gtrsim0.5\kpc$. 
This is is illustrated in Fig.~\ref{fig:vf}, which shows that the vertical variations of the net vertical force per unit mass are indeed similar in detail to those of the magnetic pressure gradient.

%-------------------------------------------------------
\begin{figure*}
    \centering    
    \includegraphics[width=0.8\textwidth]{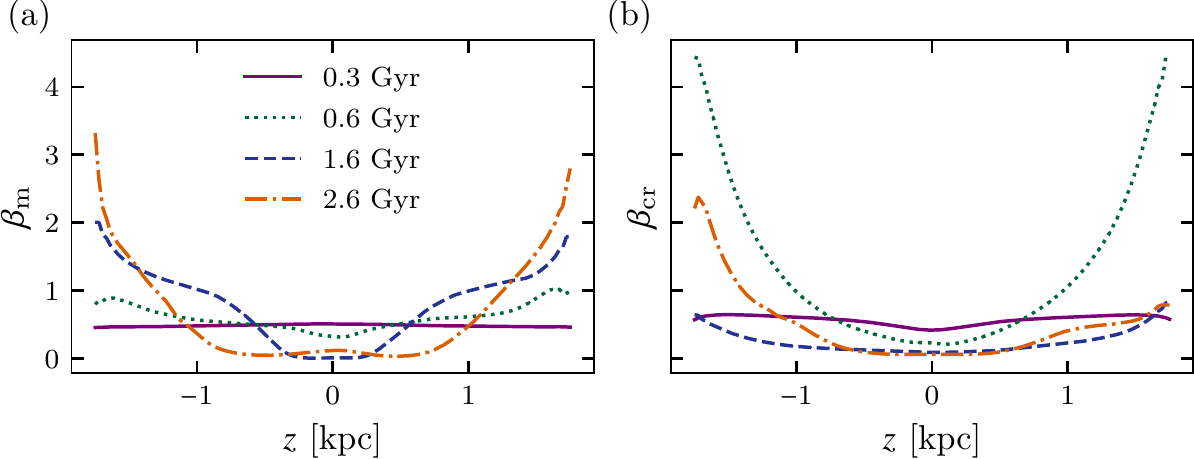}
    \caption{The distribution in $z$ of the horizontally averaged ratios of \textbf{(a)}~magnetic and \textbf{(b)}~cosmic ray pressures to the thermal pressure in Model~\SimB, $\beta\m$ and $\beta\cra$, respectively, at various times specified in the legend: the linear state, $t=0.3\Gyr$ (solid), transitional period, $t=0.6\Gyr$ (dotted), nonlinear state at $t=1.6\Gyr$ when the magnetic field reversal occurs (dashed) and a late nonlinear state, $t=2.6\Gyr$ (dash-dotted).}
    \label{fig:beta}
\end{figure*}
%---------------------------------------------------------

The magnetic and cosmic ray pressure gradients are weak because both non-thermal components of the simulated ISM are much less stratified than the thermal gas. However, their energy densities are large and they dominate over the thermal gas at $|z|\gtrsim0.5\text{--}1\kpc$. Figure~\ref{fig:beta} shows the vertical profiles of the horizontally averaged ratios of the magnetic and cosmic ray pressures to the thermal pressure, $\beta\m$ and $\beta\cra$ respectively, defined as in equation~\eqref{PmPc} but for the evolving quantities. Although each non-thermal pressure component is subdominant near the midplane at all stages of the evolution, each of them exceeds the thermal pressure at larger altitudes as soon as the instability becomes nonlinear, $t\gtrsim0.6\Gyr$. It is useful to compare Fig.~\ref{fig:beta} with Fig.~18 of \citet{SPI}: rotation somewhat reduces the magnitudes of $\beta\m$ and $\beta\cra$ at large $|z|$ but leads to the dominance of the non-thermal pressure components at smaller values of $|z|$ than in a non-rotating system, and leads to a larger contribution from cosmic rays.

%--------------------------------------------------
\section{Discussion and conclusions}\label{DC}
Differential rotation affects the nonlinear state of the Parker instability
more strongly than its linear properties. 
Without rotation, the system loses most of its magnetic field and cosmic rays as it evolves towards the steady state. 
A solid-body rotation does not change the nonlinear state significantly. 
However, differential rotation allows the system to retain better both the magnetic field and cosmic rays. The reason for that is the dynamo action (present also under the solid-body rotation but significantly enhanced by the differential rotation) which produces strong (about $2\text{--}3\muG$) large-scale magnetic field both near the midplane and at large altitudes. 
As a result, cosmic rays (governed by anisotropic diffusion) spend longer times within the system.

The systematic vertical gas flows are also affected by the rotation, which prolongs the transient outflow at a speed $|\meanh{u_z}|=7\kms$ to the time interval $0.6\lesssim t \lesssim3\Gyr$. It appears that the magnetic field contributes significantly to driving the outflow. Meanwhile, cosmic rays do not play any significant role in driving the outflow at the scales explored here, $|z|\leq1.5\kpc$: because of the large diffusivity of cosmic rays, the vertical gradient of their pressure is very small.

Another dramatic effect of the dynamo action is that it leads to a reversal of the large-scale magnetic field, in what appears to be a sign of nonlinear oscillations of the large-scale magnetic field. 
Neither the Parker instability nor the dynamo are oscillatory by themselves. We have identified the rather subtle mechanism of the reversal and argue that it is an essentially nonlinear phenomenon. 

The reversal of the large-scale magnetic field is also reflected in its spatial distribution. The reversal starts near the midplane and then the reversed magnetic field spreads to larger altitudes (see Fig.~\ref{fig:bxbyxy}e--f). As a result, the direction of the large-scale magnetic field reverses along $z$ at any given time. An arguably similar pattern of regions with the sign of the Faraday depth alternating along the direction perpendicular to the disc plane is observed in the edge-on galaxy NGC~4631 \citep{N4631}. The comparison of Figs~\ref{fig:bxbyxy}e--f and \ref{fig:bxbyxy}c--d shows that the Parker instability in a dynamo active system can produce rather complicated magnetic field structures. 
Our use of horizontal averages in Fig.~\ref{fig:bxbyxy} and elsewhere in the text conceals strong localised vertical magnetic fields typical of the magnetic buoyancy (see, e.g., Fig.~\ref{fig:3d_vi_simb}), also observed in NGC~4631. 
Because of the low gas density at kpc-scale distances from the galactic midplane, observations of the Faraday rotation produced there are difficult;
the observations of \citet{N4631} are the first of this kind, and future observation should show how widespread are such complex patterns. 
Further observational and theoretical studies of large-scale magnetic fields outside the discs of spiral galaxies promise new, unexpected insights into the dynamics of the interstellar gas and its magnetic fields. 

An unusual feature of our results, which needs further effort to be understood, is that the mean kinetic helicity of the flows driven by the Parker and magnetic buoyancy instabilities is positive in the upper half-space, $z>0$, and thus has the sign opposite to that in conventional stratified, rotating, non-magnetised systems. We note that positive kinetic helicity also occurs in some earlier studies of the mean-field dynamo action and $\alpha$-effect in magnetically-driven systems. However, this remarkable  circumstance, which can have profound --- and poorly understood --- consequences for our understanding of the nature of large-scale magnetic fields outside galactic discs, has attracted relatively little attention.

%---------------------------------------------
\section*{Acknowledgements}
We are grateful to Axel Brandenburg and Kandaswamy Subramanian for useful discussions. 
G.R.S. would like to thank the Isaac Newton Institute for Mathematical Sciences, Cambridge, for support and hospitality during the programme 'Frontiers in dynamo theory: from the Earth to the stars', where work on this paper was undertaken. This work was supported by EPSRC grant no.\ EP/R014604/1.

%%%%%%%%%%%%%%%%%%%%%%%%%%%%%%%%%%%%%%%%%%%%%%%%%%
\section*{Data Availability}
The raw data for this work were obtained from numerical simulations using the open-source PENCIL-CODE available at \url{https://github.com/pencil-code/pencil-code.git}). The derived data used for the analysis are available on request from Devika Tharakkal.

%%%%%%%%%%%%%%%%%%%% REFERENCES %%%%%%%%%%%%%%%%%%
%\bibliographystyle{mnras}
%\bibliography{Parker_rotation}

\begin{thebibliography}{}
	\makeatletter
	\relax
	\def\mn@urlcharsother{\let\do\@makeother \do\$\do\&\do\#\do\^\do\_\do\%\do\~}
	\def\mn@doi{\begingroup\mn@urlcharsother \@ifnextchar [ {\mn@doi@}
		{\mn@doi@[]}}
	\def\mn@doi@[#1]#2{\def\@tempa{#1}\ifx\@tempa\@empty \href
		{http://dx.doi.org/#2} {doi:#2}\else \href {http://dx.doi.org/#2} {#1}\fi
		\endgroup}
	\def\mn@eprint#1#2{\mn@eprint@#1:#2::\@nil}
	\def\mn@eprint@arXiv#1{\href {http://arxiv.org/abs/#1} {{\tt arXiv:#1}}}
	\def\mn@eprint@dblp#1{\href {http://dblp.uni-trier.de/rec/bibtex/#1.xml}
		{dblp:#1}}
	\def\mn@eprint@#1:#2:#3:#4\@nil{\def\@tempa {#1}\def\@tempb {#2}\def\@tempc
		{#3}\ifx \@tempc \@empty \let \@tempc \@tempb \let \@tempb \@tempa \fi \ifx
		\@tempb \@empty \def\@tempb {arXiv}\fi \@ifundefined
		{mn@eprint@\@tempb}{\@tempb:\@tempc}{\expandafter \expandafter \csname
			mn@eprint@\@tempb\endcsname \expandafter{\@tempc}}}
	
	\bibitem[\protect\citeauthoryear{{Brandenburg} \& {Schmitt}}{{Brandenburg} \&
		{Schmitt}}{1998}]{BrSc98}
	{Brandenburg} A.,  {Schmitt} D.,  1998, \aap, \href
	{https://ui.adsabs.harvard.edu/abs/1998A&A...338L..55B} {338, L55}
	
	\bibitem[\protect\citeauthoryear{{Brandenburg} \& {Sokoloff}}{{Brandenburg} \&
		{Sokoloff}}{2002}]{BrSo02}
	{Brandenburg} A.,  {Sokoloff} D.,  2002, \mn@doi [Geophys.\ Astrophys.\ Fluid
	Dyn.] {10.1080/03091920290032974}, \href
	{https://ui.adsabs.harvard.edu/abs/2002GApFD..96..319B} {96, 319}
	
	\bibitem[\protect\citeauthoryear{{Brandenburg}, {Nordlund}, {Stein}  \&
		{Torkelsson}}{{Brandenburg} et~al.}{1995}]{Brandenburg1995}
	{Brandenburg} A.,  {Nordlund} A.,  {Stein} R.~F.,   {Torkelsson} U.,  1995,
	\mn@doi [\apj] {10.1086/175831}, \href
	{https://ui.adsabs.harvard.edu/abs/1995ApJ...446..741B} {446, 741}
	
	\bibitem[\protect\citeauthoryear{{Foglizzo} \& {Tagger}}{{Foglizzo} \&
		{Tagger}}{1994}]{FogTag1994}
	{Foglizzo} T.,  {Tagger} M.,  1994, \aap, \href
	{https://ui.adsabs.harvard.edu/abs/1994A&A...287..297F} {287, 297}
	
	\bibitem[\protect\citeauthoryear{{Foglizzo} \& {Tagger}}{{Foglizzo} \&
		{Tagger}}{1995}]{FogTag1995}
	{Foglizzo} T.,  {Tagger} M.,  1995, \aap, \href
	{https://ui.adsabs.harvard.edu/abs/1995A&A...301..293F} {301, 293}
	
	\bibitem[\protect\citeauthoryear{{Hanasz}}{{Hanasz}}{1997}]{Hanasz1997b}
	{Hanasz} M.,  1997, \aap, \href
	{https://ui.adsabs.harvard.edu/abs/1997A&A...327..813H} {327, 813}
	
	\bibitem[\protect\citeauthoryear{{Hanasz} \& {Lesch}}{{Hanasz} \&
		{Lesch}}{1997}]{Hanasz1997a}
	{Hanasz} M.,  {Lesch} H.,  1997, \aap, \href
	{https://ui.adsabs.harvard.edu/abs/1997A&A...321.1007H} {321, 1007}
	
	\bibitem[\protect\citeauthoryear{{Hanasz} \& {Lesch}}{{Hanasz} \&
		{Lesch}}{1998}]{Hanasz1998}
	{Hanasz} M.,  {Lesch} H.,  1998, \aap, \href
	{https://ui.adsabs.harvard.edu/abs/1998A&A...332...77H} {332, 77}
	
	\bibitem[\protect\citeauthoryear{{Hanasz}, {Kowal}, {Otmianowska-Mazur}  \&
		{Lesch}}{{Hanasz} et~al.}{2004}]{Hanasz2004}
	{Hanasz} M.,  {Kowal} G.,  {Otmianowska-Mazur} K.,   {Lesch} H.,  2004, \mn@doi
	[\apjl] {10.1086/420697}, \href
	{https://ui.adsabs.harvard.edu/abs/2004ApJ...605L..33H} {605, L33}
	
	\bibitem[\protect\citeauthoryear{{Kowal}, {Hanasz}  \&
		{Otmianowska-Mazur}}{{Kowal} et~al.}{2003}]{KoHaOt2003}
	{Kowal} G.,  {Hanasz} M.,   {Otmianowska-Mazur} K.,  2003, \mn@doi [\aap]
	{10.1051/0004-6361:20030556}, \href
	{https://ui.adsabs.harvard.edu/abs/2003A&A...404..533K} {404, 533}
	
	\bibitem[\protect\citeauthoryear{{Kuijken} \& {Gilmore}}{{Kuijken} \&
		{Gilmore}}{1989}]{kg89}
	{Kuijken} K.,  {Gilmore} G.,  1989, \mn@doi [\mnras] {10.1093/mnras/239.2.571},
	\href {https://ui.adsabs.harvard.edu/abs/1989MNRAS.239..571K} {239, 571}
	
	\bibitem[\protect\citeauthoryear{{Machida}, {Nakamura}, {Kudoh}, {Akahori},
		{Sofue}  \& {Matsumoto}}{{Machida} et~al.}{2013}]{Machida2013}
	{Machida} M.,  {Nakamura} K.~E.,  {Kudoh} T.,  {Akahori} T.,  {Sofue} Y.,
	{Matsumoto} R.,  2013, \mn@doi [\apj] {10.1088/0004-637X/764/1/81}, \href
	{https://ui.adsabs.harvard.edu/abs/2013ApJ...764...81M} {764, 81}
	
	\bibitem[\protect\citeauthoryear{Matsuzaki, Matsumoto, Tajima  \&
		Shibata}{Matsuzaki et~al.}{1998}]{TMatRMat1998}
	Matsuzaki T.,  Matsumoto R.,  Tajima T.,   Shibata K.,  1998, in Watanabe T.,
	Kosugi T.,   Sterling A.~C.,  eds, Observational Plasma Astrophysics: Five
	Years of Yohkoh and Beyond. Springer Netherlands, Dordrecht, pp 321--324,
	\mn@doi{10.1007/978-94-011-5220-4_52}
	
	\bibitem[\protect\citeauthoryear{{Mora-Partiarroyo} et~al.,}{{Mora-Partiarroyo}
		et~al.}{2019}]{N4631}
	{Mora-Partiarroyo} S.~C.,  et~al., 2019, \mn@doi [\aap]
	{10.1051/0004-6361/201935961}, \href
	{https://ui.adsabs.harvard.edu/abs/2019A&A...632A..11M} {632, A11}
	
	\bibitem[\protect\citeauthoryear{{Moss}, {Shukurov}  \& {Sokoloff}}{{Moss}
		et~al.}{1999}]{MSS99}
	{Moss} D.,  {Shukurov} A.,   {Sokoloff} D.,  1999, \aap, \href
	{https://ui.adsabs.harvard.edu/abs/1999A&A...343..120M} {343, 120}
	
	\bibitem[\protect\citeauthoryear{{Oishi} \& {Mac Low}}{{Oishi} \& {Mac
			Low}}{2011}]{Oishi2011}
	{Oishi} J.~S.,  {Mac Low} M.-M.,  2011, \mn@doi [\apj]
	{10.1088/0004-637X/740/1/18}, \href
	{https://ui.adsabs.harvard.edu/abs/2011ApJ...740...18O} {740, 17}
	
	\bibitem[\protect\citeauthoryear{{Ruzmaikin}, {Sokoloff}  \&
		{Turchaninov}}{{Ruzmaikin} et~al.}{1980}]{RST80}
	{Ruzmaikin} A.~A.,  {Sokoloff} D.~D.,   {Turchaninov} V.~L.,  1980, \sovast,
	\href {https://ui.adsabs.harvard.edu/abs/1980SvA....24..182R} {24, 182}
	
	\bibitem[\protect\citeauthoryear{{Seta}, {Shukurov}, {Wood}, {Bushby}  \&
		{Snodin}}{{Seta} et~al.}{2018}]{SSWBS18}
	{Seta} A.,  {Shukurov} A.,  {Wood} T.~S.,  {Bushby} P.~J.,   {Snodin} A.~P.,
	2018, \mn@doi [\mnras] {10.1093/mnras/stx2606}, \href
	{https://ui.adsabs.harvard.edu/abs/2018MNRAS.473.4544S} {473, 4544}
	
	\bibitem[\protect\citeauthoryear{{Shu}}{{Shu}}{1974}]{Shu1974}
	{Shu} F.~H.,  1974, \aap, \href
	{https://ui.adsabs.harvard.edu/abs/1974A&A....33...55S} {33, 55}
	
	\bibitem[\protect\citeauthoryear{{Shukurov} \& {Subramanian}}{{Shukurov} \&
		{Subramanian}}{2021}]{ShSu22}
	{Shukurov} A.,  {Subramanian} K.,  2021, Astrophysical Magnetic Fields: From
	Galaxies to the Early Universe.
	Cambridge University Press, Cambridge, \mn@doi{10.1017/9781139046657}
	
	\bibitem[\protect\citeauthoryear{{Tharakkal}, {Snodin}, {Sarson}  \&
		{Shukurov}}{{Tharakkal} et~al.}{2022b}]{TSSS22}
	{Tharakkal} D.,  {Snodin} A.~P.,  {Sarson} G.~R.,   {Shukurov} A.,  2022b,
	arXiv:2205.01986, \href
	{https://ui.adsabs.harvard.edu/abs/2022arXiv220501986T} {pp 1--19}
	
	\bibitem[\protect\citeauthoryear{{Tharakkal}, {Shukurov}, {Gent}, {Sarson},
		{Snodin}  \& {Rodrigues}}{{Tharakkal} et~al.}{2022a}]{SPI}
	{Tharakkal} D.,  {Shukurov} A.,  {Gent} F.~A.,  {Sarson} G.~R.,  {Snodin}
	A.~P.,   {Rodrigues} L. F.~S.,  2022a, arXiv:2212.03215, \href
	{https://ui.adsabs.harvard.edu/abs/2022arXiv221203215T} {pp 1--18}
	
	\bibitem[\protect\citeauthoryear{{Thelen}}{{Thelen}}{2000a}]{The00a}
	{Thelen} J.~C.,  2000a, \mn@doi [\mnras] {10.1046/j.1365-8711.2000.03419.x},
	\href {https://ui.adsabs.harvard.edu/abs/2000MNRAS.315..155T} {315, 155}
	
	\bibitem[\protect\citeauthoryear{{Thelen}}{{Thelen}}{2000b}]{The00b}
	{Thelen} J.~C.,  2000b, \mn@doi [\mnras] {10.1046/j.1365-8711.2000.03420.x},
	\href {https://ui.adsabs.harvard.edu/abs/2000MNRAS.315..165T} {315, 165}
	
	\bibitem[\protect\citeauthoryear{{Zweibel} \& {Kulsrud}}{{Zweibel} \&
		{Kulsrud}}{1975}]{ZwKu1975}
	{Zweibel} E.~G.,  {Kulsrud} R.~M.,  1975, \mn@doi [\apj] {10.1086/153858},
	\href {https://ui.adsabs.harvard.edu/abs/1975ApJ...201...63Z} {201, 63}
	
	\makeatother
\end{thebibliography}

%%%%%%%%%%%%%%%%%%%%%%%%%%%%%%%%%%%%%%%%%%%%%%%%%%

%%%%%%%%%%%%%%%%% APPENDICES %%%%%%%%%%%%%%%%%%%%%
%%%%%%%%%%%%%%%%%%%%%%%%%%%%%%%%%%%%%%%%%%%%%%%%%%

% Don't change these lines
\bsp	% typesetting comment
\label{lastpage}
\end{document}